\documentclass[aps,prl,superscriptaddress,twocolumn]{revtex4-1}
\usepackage{bbm}
\usepackage{mathrsfs}
\usepackage{amsmath}
\usepackage{amsfonts}
\usepackage[colorlinks=true,citecolor=blue,anchorcolor=blue]{hyperref}
\usepackage{graphicx,epstopdf}
\usepackage{subfigure}
\usepackage{epsfig}
\usepackage{dcolumn}
\usepackage{bm}
\usepackage{color}
\usepackage{natbib}
\usepackage{amssymb}
\usepackage{xcolor}
\usepackage{braket}

\begin{document}
\title{Remote generation of magnon Schr\"{o}dinger cat state via magnon-photon entanglement}

\author{Feng-Xiao Sun}
\affiliation{State Key Laboratory for Mesoscopic Physics, School of Physics, Frontiers Science Center for Nano-optoelectronics, $\&$ Collaborative
Innovation Center of Quantum Matter, Peking University, Beijing 100871, China}
\affiliation{Collaborative Innovation Center of Extreme Optics, Shanxi University, Taiyuan, Shanxi 030006, China}
\author{Sha-Sha Zheng}
\affiliation{State Key Laboratory for Mesoscopic Physics, School of Physics, Frontiers Science Center for Nano-optoelectronics, $\&$ Collaborative
Innovation Center of Quantum Matter, Peking University, Beijing 100871, China}
\affiliation{Collaborative Innovation Center of Extreme Optics, Shanxi University, Taiyuan, Shanxi 030006, China}
\author{Yang Xiao}
\affiliation{Department of Applied Physics, Nanjing University of Aeronautics and Astronautics, Nanjing 210016, China}
\author{Qihuang Gong}
\affiliation{State Key Laboratory for Mesoscopic Physics, School of Physics, Frontiers Science Center for Nano-optoelectronics, $\&$ Collaborative
Innovation Center of Quantum Matter, Peking University, Beijing 100871, China}
\affiliation{Collaborative Innovation Center of Extreme Optics, Shanxi University, Taiyuan, Shanxi 030006, China}
\affiliation{Peking University Yangtze Delta Institute of Optoelectronics, Nantong 226010, Jiangsu, China}
\author{Qiongyi He}
\email{qiongyihe@pku.edu.cn}
\affiliation{State Key Laboratory for Mesoscopic Physics, School of Physics, Frontiers Science Center for Nano-optoelectronics, $\&$ Collaborative
Innovation Center of Quantum Matter, Peking University, Beijing 100871, China}
\affiliation{Collaborative Innovation Center of Extreme Optics, Shanxi University, Taiyuan, Shanxi 030006, China}
\affiliation{Peking University Yangtze Delta Institute of Optoelectronics, Nantong 226010, Jiangsu, China}
\author{Ke Xia}
\email{kexia@csrc.ac.cn}
\affiliation{Beijing Computational Science Research Center, Beijing 100193, China}

\begin{abstract}
	Magnon cat state represents a macroscopic quantum superposition of collective magnetic excitations of large number spins that not only provides fundamental tests of macroscopic quantum effects but also finds applications in quantum metrology and quantum computation. In particular, remote generation and manipulation of Schr\"{o}dinger cat states are particularly interesting for the development of long-distance and large-scale quantum information processing. Here, we propose an approach to remotely prepare magnon even/odd cat states by performing local non-Gaussian operations on the optical mode that is entangled with magnon mode through pulsed optomagnonic interaction. By evaluating key properties of the resulting cat states, we show that for experimentally feasible parameters they are generated with both high fidelity and nonclassicality, and with a size large enough to be useful for quantum technologies. Furthermore, the effects of experimental imperfections such as the error of projective measurements and dark count when performing single-photon operations have been discussed, where the lifetime of the created magnon cat states is expected to be $t\sim1\,\mu$s.

\end{abstract}
\maketitle

\textit{Introduction.--} 
          Magnons are collective spin excitations of magnetic materials that due to their long lifetime and high spin density represent ideal candidates for studies of macroscopic quantum effects. These quasiparticles also open new possibilities for quantum technologies as carriers of quantum information, since they are widely frequency-tunable and can be coherently coupled to different degrees of freedom~\cite{lachance2019hybrid}, such as microwave photons~\cite{soykal2010strong,huebl2013high,zhang2014strongly,tabuchi2014hybridizing,cao2015exchange,bai2015spin}, optical photons~\cite{osada2016cavity,haigh2016triple,zhang2016optomagnonic,osada2018brillouin}, superconducting qubits~\cite{tabuchi2015coherent,lachance2017resolving,wolski2020dissipation}, as well as phonons~\cite{zhang2016cavity,kikkawa2016magnon,bozhko2020magnon}.  Moreover, remarkable experiments, e.g. the observation of magnon Bose-Einstein condensation~\cite{demokritov2006bose} raised considerable interest in the study of fundamental quantum features of magnons such as  entanglement~\cite{li2018magnon,yuan2020enhancement,*yuan2020steady,lachance2020entanglement} and squeezing~\cite{kamra2016super,li2019squeezed,yuan2021unconventional} in macroscopic systems.

Of particular interest is the preparation of  {\it Schr\"{o}dinger cat state}~\cite{schrodinger1935gegenwartige} in such massive systems, i.e.~the quantum superposition of two macroscopically distinguishable magnon coherent states. These states are not only interesting for testing the fundamentals of quantum mechanics such as the quantum-to-classical transition~\cite{wineland2013nobel,*arndt2014testing,*li2016quantum}, but also represent a valuable resource for fault-tolerant quantum computing~\cite{cochrane1999macroscopically,lund2008fault,ofek2016extending} and quantum metrology~\cite{zurek2001sub,joo2011quantum}. The size of cat state is of crucial importance for any desired applications~\cite{cochrane1999macroscopically,lund2008fault,ofek2016extending,zurek2001sub,joo2011quantum}. For well-studied ferromagnetic insulators, e.g. Yttrium Iron Garnet (YIG) spheres, their ultra-low magnetic dissipation and high spin density $\sim10^{18}/\text{mm}^3$~\cite{zhang2014strongly,tabuchi2014hybridizing} make realizing a cat state of a macroscopic number of spins feasible. 

A well established method in quantum optics consists of adding or subtracting single photon on a squeezed vacuum state, which results in a cat state locally at the site where the operations are implemented~\cite{dakna1997generating,ourjoumtsev2006generating}. This approach has been recently investigated for the preparation of a magnon cat state in an ideal system without decoherence~\cite{sharma2021spin}. As the coupling to environment is inevitable in all practical situations, it is crucial to explore efficient ways to create magnon cat states in realistic conditions~\cite{rahman2019large} and evaluate their nonclassical qualities by taking damping processes into account.

Motivated by the quest for long-distance quantum network with long-lived magnons, in this letter we propose a different scheme aiming to $remotely$ generate and manipulate a magnon cat state. Specifically, we employ a pulsed optomagnonic coupling scheme to prepare magnon-photon entanglement and then transmit the light to a distant position. By performing appropriate single-photon operations and projective measurement on the optical mode, the magnon collapses to an odd/even cat state. We examine key figures of merit for the produced cat state, such as Wigner negativity~\cite{kenfack2004negativity}, macroscopic quantum superposition~\cite{lee2011quantification}, and fidelity~\cite{jozsa1994fidelity}. A relatively large size of coherent components $|\alpha|^2 \approx 1.44 ~(3.42)$ with high fidelity is achieved for odd (even) cat. The parameters chosen in the present scheme are all in the accessible range with the state-of-the-art experimental techniques. Besides, the imperfections of projective measurement and the dark counts in single-photon operations are considered, where the reduction of qualities is negligible, with the lifetime $t\sim1\,\mu$s.
\begin{figure}[tb]
	\centering
	\includegraphics[width=0.46\textwidth]{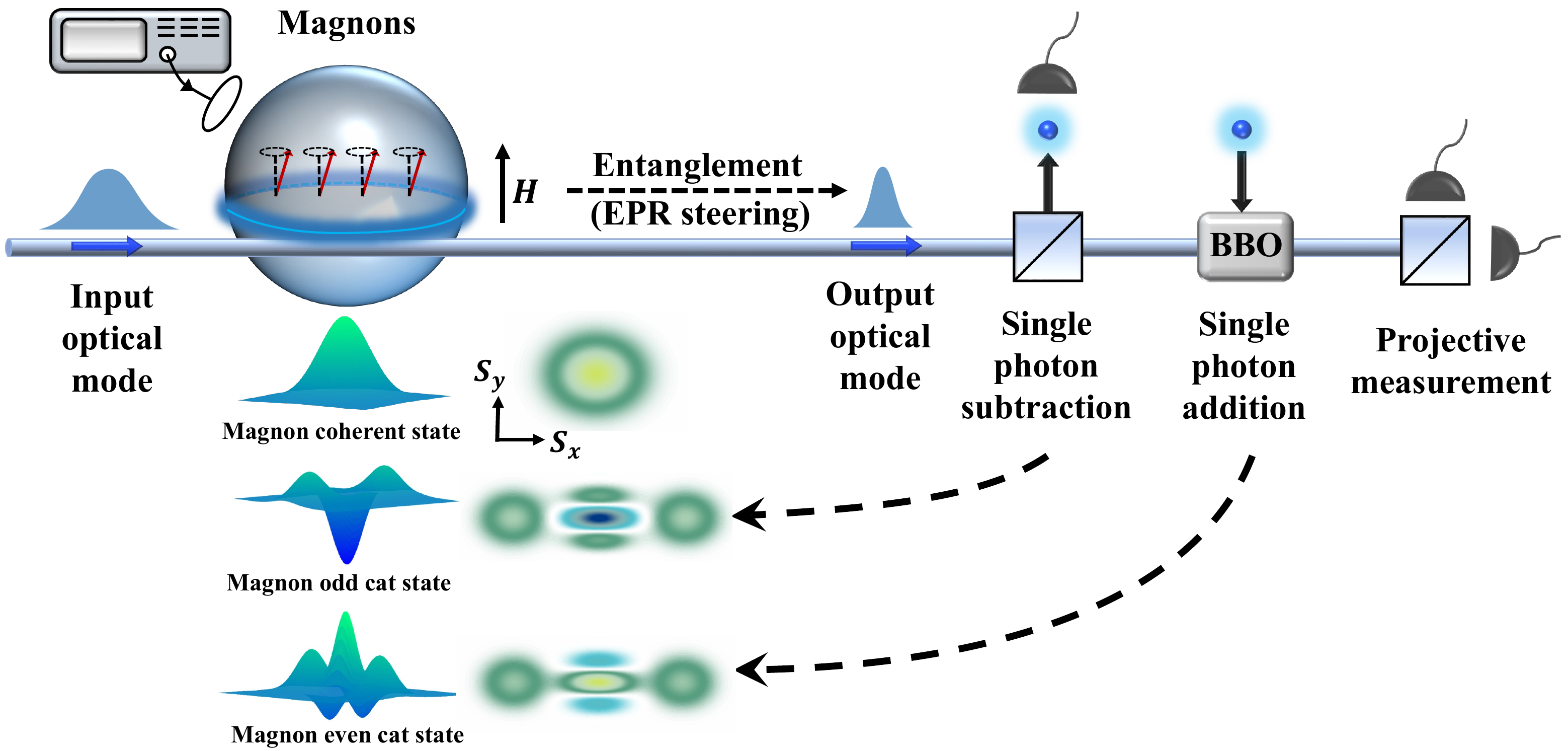}
	\caption{Schematic diagram for the remote generation of a magnon cat state. The magnons are prepared with an external magnetic field $H$, and they couple with the optical mode through optomagnonic interaction that results in magnon-photon entanglement. Then, by applying a single-photon operation on the optical mode, the initial magnon coherent state turns into a non-Gaussian state. After a projective measurement on the optical mode, the magnon's state collapses into a Schr\"{o}dinger cat state. Odd (even) magnon cat state can be remotely generated by performing a single-photon subtraction (followed by a single-photon addition).}
	\label{fig:schematic}
\end{figure}

\textit{Pulsed cavity optomagnonic model.--} 
        Cavity optomagnonics has recently attracted great interests as a new platform for studying quantum entanglement~\cite{yu2020magnetostrictively} and for quantum technologies such as quantum memory~\cite{zhang2015magnon}. Contrary to optomechanical systems, where the mechanical oscillator's frequency is fixed, the frequency of magnon mode can be continuously tuned in the range of microwave frequency by varying the external magnetic field~\cite{soykal2010strong,huebl2013high,zhang2014strongly,tabuchi2014hybridizing,cao2015exchange,bai2015spin,chumak2015magnon}.  The magnon can also be coupled to different degrees of freedom~\cite{lachance2019hybrid} due to various magnetic interactions. These advantages make flexible control on optomagnonics possible and facilitate the coupling to other quantum systems. The higher frequency of magnon in the GHz range also benefits for ground-state cooling than the MHz frequency of phonon modes in optomechanics.

We consider a YIG sphere impinged by a light beam propagating tangentially to the equator, which introduces the coupling between the optical whispering gallery mode and the magnon mode of YIG sphere, Fig.~\ref{fig:schematic}. The system is prepared at low temperature $T<0.2\,$K with negligible thermal occupation~\cite{tabuchi2014hybridizing,o2010quantum}, where the details and the effects of thermal occupation are discussed in~\cite{supplementary}. The magnon mode is initially in a coherent spin state with isotropic uncertainty of the macrospin components $S_x$ and $S_y$ due to quantum fluctuations, which is associated to a positive Gaussian Wigner quasiprobability distribution~\cite{scully2012quantum}. Then, we exploit a pulsed optomagnonic interaction to prepare magnon-photon entanglement that is strong enough to exhibit Einstein-Podolsky-Rosen (EPR) steering~\cite{reid1989demonstration,wiseman2007steering}. The optomagnonic interaction energy is expressed by $E_{OM}=\frac{1}{4} \int d\mathbf{r}\, \mathbf{E}^*(\mathbf{r}) \varepsilon(\mathbf{M}) \mathbf{E}(\mathbf{r})$, where $\mathbf{E}(\mathbf{r})$ is the electromagnetic field and $\varepsilon(\mathbf{M})$ is the magnetization-dependent dielectric constant tensor. Such optomagnonic coupling has been demonstrated in recent experiments~\cite{osada2016cavity,zhang2016optomagnonic,haigh2016triple,osada2018brillouin}. In the limit of small spin excitations, the Holstein-Primakoff approximation allows us to describe the magnon mode as a bosonic mode with annihilation operator $m\propto S_x+iS_y$~\cite{holstein1940field,supplementary}. Here, we focus on the uniform magnetostatic mode, i.e.~Kittel mode~\cite{osada2016cavity,haigh2016triple}, while surface magnetostatic modes present similar physics in the study of cavity optomagnonics~\cite{sharma2017light,*sharma2018optical}.

The effective interaction Hamiltonian for the optomagnonic system in a frame rotating at the laser frequency $\omega_p$ is then~\cite{supplementary}
\begin{equation}
	H=\hbar\omega_{m}m^{\dagger}m+\hbar\Delta c^{\dagger}c+\hbar g_{0}c^{\dagger}c(m^{\dagger}+m),
	\label{Hamiltonian_optomagnon}
\end{equation}
where $c$ is the annihilation operator for the optical mode, $g_0$ is the optomagnonic coupling strength, $\omega_{m},~\omega_{c}$ are the frequencies of the magnon and optical modes, respectively, and $\Delta=\omega_{c}-\omega_{p}$ is the detuning of the cavity with respect to the pumped laser.

\begin{figure}[b]
	\centering
	\includegraphics[width=0.3\textwidth]{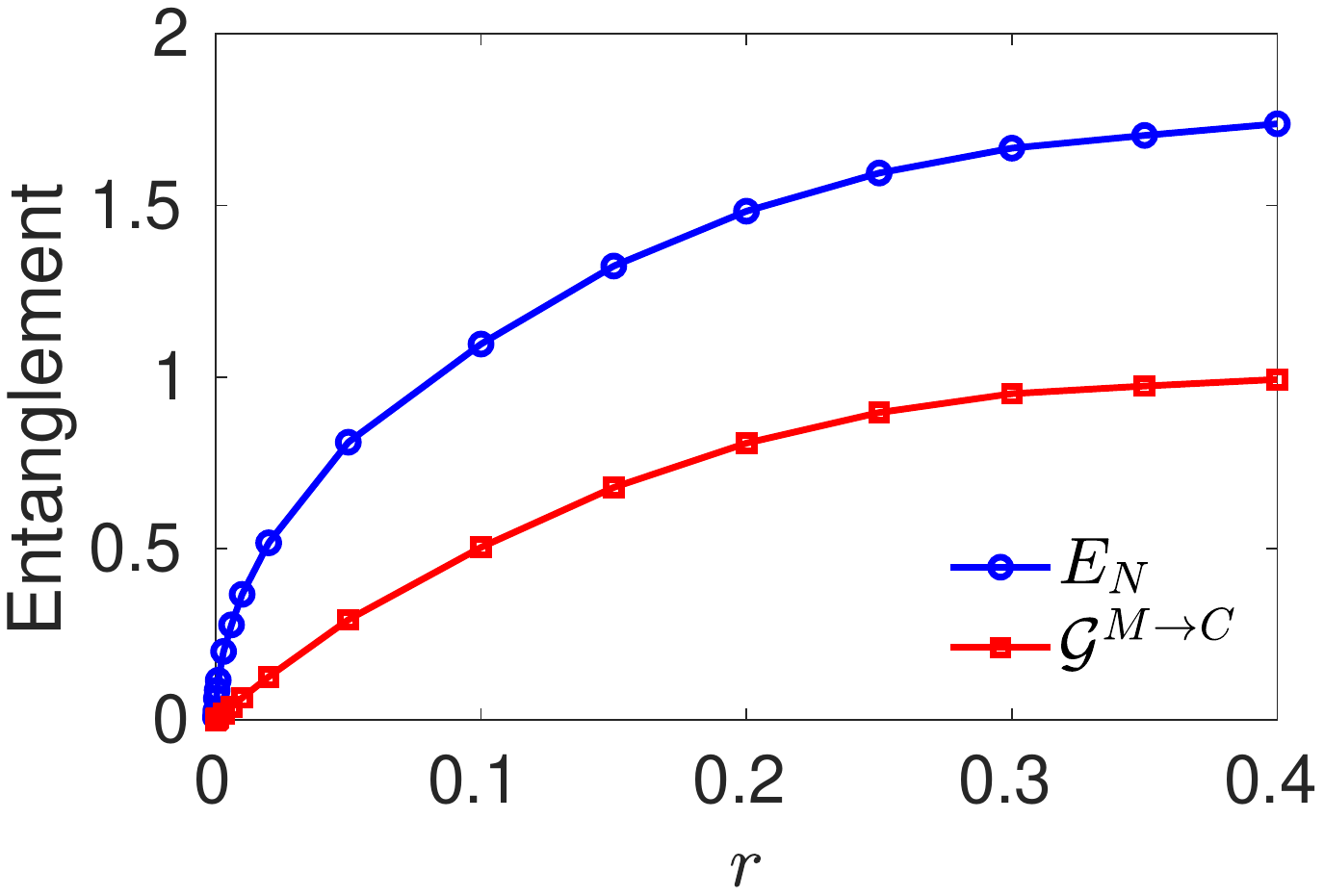}
	\caption{Entanglement $E_N$ and EPR steering from magnon mode to optical mode $\mathcal{G}^{M\to C}$ as functions of effective squeezing parameter $r=G\tau$. Other parameters are $g/2\pi=5\,$MHz, $\kappa/2\pi=100\,$MHz, $\gamma/2\pi=0.1\,$MHz.}
	\label{fig:EPR}
\end{figure}
\begin{figure*}[t]
	\centering
	\includegraphics[width=0.96\textwidth]{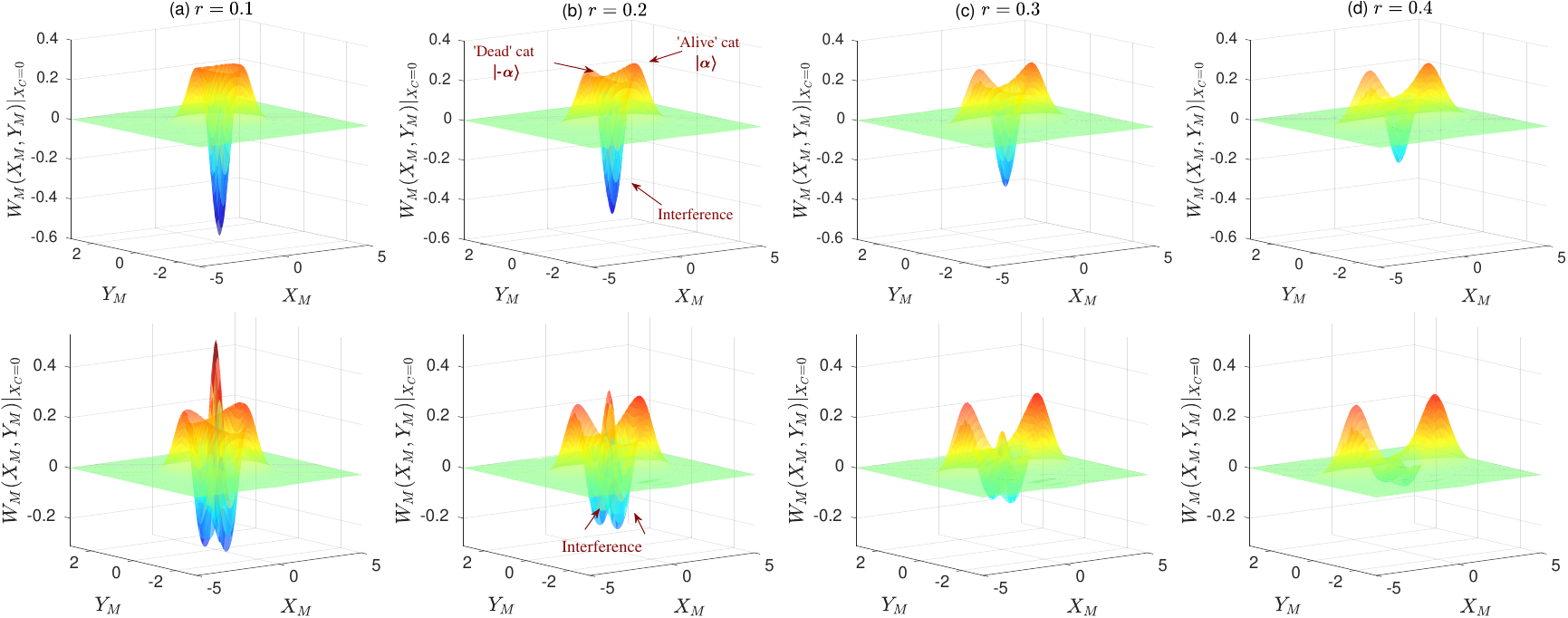}
	\caption{Wigner functions of the transient magnon cat states. The effective squeezing parameters $r=G\tau=0.1$ (a),~$0.2$ (b),~$0.3$ (c),~$0.4$ (d), where $G=g^2/\kappa-\gamma$ and $\tau$ is the pulse time. Upper row: Wigner function of the projected magnon mode $M$ for the outcome $X_C=0$ after subtracting single photon from the optical mode $C$, displaying features of an odd cat state. Lower row: The transient even cat states by performing a single-photon subtraction and single-photon addition in sequence on the optical mode before further projective measurement. The parameters are the same as in Fig.~\ref{fig:EPR}.}
	\label{fig:cats_in_time}
\end{figure*}

We pump the cavity mode with a blue-detuned pulse with $\Delta=-\omega_{m}$, which results in a two-mode squeezing type of interaction that entangles the magnon mode with the optical mode~\cite{palomaki2013entangling}. This becomes evident by linearizing the quantum fluctuations of the optical mode around the mean value, i.e.~substituting $c \rightarrow \beta + c$, and then applying the rotating wave approximation. The Hamiltonian is then dominated by,
\begin{equation}
	H_{\text{RWA}}=\hbar g (mc+m^{\dagger}c^{\dagger}),
	\label{Hamiltonian_RWA}
\end{equation}
where $g=g_0|\beta|$. The details of this derivation, together with the quantum Langevin equations for the modes $m$ and $c$, are provided in Section SI of~\cite{supplementary}. In recent experiments, the effective optomagnonic coupling has already reached $g/2\pi\sim10\,$kHz~\cite{zhang2016optomagnonic}. It could be further enhanced up to $g/2\pi\sim10\,$MHz by decreasing the diameter of YIG sphere to $~1\,\mu$m (limited by the optical wavelength) that yields a total spin number of $\sim10^{10}$ and thus remarkably large $g_0/2\pi\approx0.1\,$MHz, together with enhancing the cavity light amplitude $|\beta|\sim100$ for a typical microcavity~\cite{kusminskiy2016coupled}. As for the inevitable decays, the typical value for the optical cavity damping is $\kappa/2\pi\sim100\,$MHz~\cite{zhang2016optomagnonic,haigh2016triple}, and for YIG ultra-low damping $\gamma/2\pi\sim0.1\,$MHz is promising to be realized since the magnon linewidth $\gamma/2\pi\sim 0.6\,$MHz with temperature $0.1\,\text{K}<T<1\,$K has been reported~\cite{tabuchi2014hybridizing}. By further suppressing the impurity-magnon and other scattering, one expects to achieve a lower value of damping $\gamma$ for high quality YIG samples. We discussed the results with larger damping $\gamma$ in~\cite{supplementary} and found that magnon cat states can be prepared remotely by further improving the coupling $g$.
Higher-order magnon modes are ignored since a low magnon damping indicates a weak magnon-magnon scattering~\cite{tabuchi2014hybridizing}.

From the input-output relation, the cavity output field mode can be expressed as $c_{out}=c_{in}+\sqrt{2\kappa}c$. Since the optical mode is pumped by pulse of finite duration, it is convenient to introduce normalized temporal modes for the output optical field and the magnon~\cite{vanner2011pulsed,hofer2011quantum}, namely
\begin{align}
	C_{out}&=\sqrt{\frac{2G}{e^{2r}-1}}\int_{0}^{\tau}e^{Gt}c_{out}(t)dt ,\quad M_{out}=m(\tau).
	\label{output_mode}
\end{align}
Here, $\tau$ is the pulse duration, $r = G\tau$ is the effective squeezing parameter with $G = g^2/\kappa - \gamma$  (see details in Section SI of~\cite{supplementary}). 

\textit{The magnon-photon entanglement and EPR steering.--}
       We now investigate the quantum correlations between the optical $(C)$ and the magnon $(M)$ modes.  The quadratures are defined as $X_{O}=(O_{out}+O_{out}^{\dagger})/\sqrt{2}$, $Y_{O}=(O_{out}-O_{out}^{\dagger})/\sqrt{2}i$, with $O_{out}=C_{out},M_{out}$. Here, $X_M$ and $Y_M$ correspond to $S_x$ and $S_y$ after the Holstein-Primakoff transformation. By defining the covariance matrix as $V=\{V_C,~V_{CM};~V_{CM}^{T},~V_M\}$, where each element is a $2\times2$ block matrix with components $V_{ij}=\langle u_{i}u_{j}+u_{j}u_{i} \rangle/2$ ($i,j=1,2,3,4$), $u=(X_{C},Y_{C},X_{M},Y_{M})^{T}$, the magnon-photon entanglement can be quantified in terms of the logarithmic negativity~\cite{adesso2004extremal}, $E_{N}=\text{max}\{0,-\ln2\eta^{-}\}$ where $\eta^{-}=\sqrt{\Sigma(V)-[\Sigma(V)^{2}-4\text{det}V]^{1/2}}/\sqrt{2}$ and $\Sigma(V)=\text{det}V_{C}+\text{det}V_{M}-2\text{det}V_{CM}$. Furthermore, EPR steering from the magnon to the optical mode is given by $\mathcal{G}^{M\to C}=\text{max}\{0,\mathcal{S}(2V_{M})-\mathcal{S}(2V)\}$~\cite{kogias2015quantification}, where $\mathcal{S}(V)=\frac{1}{2}\ln\text{det}V$ is the R\'enyi-2 entropy.  In Fig.~\ref{fig:EPR} we illustrate the resulting entanglement and EPR steering as functions of $r$, and conclude that both these quantum correlations are present whenever $r>0$. 

EPR steering has recently been demonstrated as a crucial resource for the remote generation of Wigner negativity on the steering mode by performing single-photon subtraction on the steered mode~\cite{walschaers2020remote}. For our purposes, since cat states are typical non-Gaussian states with negative Wigner function, observing EPR steering will allow for their remote preparation in magnon modes, as we will show in the following.

\begin{figure}
	\centering
	\includegraphics[width=0.49\textwidth]{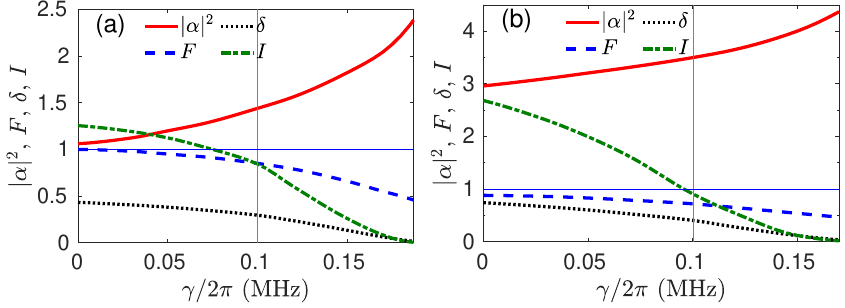}
	\caption{Cat size $|\alpha|^2$ (red solid), fidelity $F$ (blue dashed), Wigner negativity $\delta$ (black dots), and macroscopic quantum superposition $I$ (green dash-dotted) of (a) the odd cat states and (b) the even cat states remotely created varying with $\gamma$ at $r=G\tau=0.2\,$ with $g/2\pi=5\,$MHz and $\kappa/2\pi=100\,$MHz.}
	\label{fig:F-r}
\end{figure}

\textit{Transient magnon cat state.--}
    After interaction with the YIG sphere resulting in photon-magnon entanglement, the cavity output field can be transmitted to a distant site. By subtracting a photon on the output optical field~\cite{parigi2007probing,ra2020non}, the magnon mode immediately collapses into a non-Gaussian state and then becomes an odd cat state after a projective measurement on the optical mode with outcome $X_C=0$ (the cat taken the projective measurement imperfection $\epsilon=0.1$ into account, i.e.~$|X_C|\le\epsilon$, has been shown in Section SIV of~\cite{supplementary}). From the Wigner functions in the upper row of Fig.~\ref{fig:cats_in_time}, we observe two distinct peaks in $X_M$, revealing the quantum superposition phenomena where two contradictory outcomes of $X_M$ are simultaneously possible. The interference fringes between the `Dead' and `Alive' states occur because of quantum coherence. The peaks become more separated with longer optical pulse duration $\tau$ (and thus larger effective squeezing parameter $r=G\tau$), representing larger size of created cats. Note that Holstein-Primakoff approximation holds well for the case considered here, since we will show in Fig.~\ref{fig:F-r} that the number of magnons $\langle m^\dagger m\rangle\cong|\alpha|^2$ is much smaller than the total spin number of the YIG sphere.
    
Moreover, if after photon-subtraction a single-photon addition~\cite{parigi2007probing,ra2020non,zavatta2004quantum} is performed on the optical mode, the magnon will collapse to an even cat state. The difference in parity between even and odd cat states results in positive and negative amplitude at the symmetry center in the Wigner functions~\cite{supplementary}, as drawn in the lower row of Fig.~\ref{fig:cats_in_time}. In this way, we are able to prepare different magnon cat states by choosing suitable single-photon operations. 
Meanwhile, the interference patterns become less apparent from Fig.~\ref{fig:cats_in_time} (a) to (d), which results from the decoherence in the magnon and photon modes. This indicates that, due to the interactions with the environment, the magnon state will ultimately evolve from the coherent superposition of two components into a statistical (classical) mixture of them~\cite{zurek2006decoherence}.

The quality of the transient magnon cat states is quantified by the following figures of merit. First, the fidelity between the prepared state $\rho$ and an ideal cat state $|\psi_{\pm}\rangle=\mathcal{N}_{\pm}(|\alpha\rangle\pm|-\alpha\rangle)$ (normalized parameters $\mathcal{N}_{\pm}=[2(1\pm e^{-2|\alpha|^2})]^{-1/2}$), which is defined as $F=\langle \psi_{\pm}|\rho|\psi_{\pm} \rangle$~\cite{jozsa1994fidelity}. Then, the size of the cat $|\alpha|^2$, where a large value of $|\alpha|^2$ corresponds to a small overlap between the two coherent states as required for error reduction in the distinguishability of the two peaks~\cite{lund2008fault}. Wigner negativity $\delta$~\cite{kenfack2004negativity} and macroscopic quantum superposition $I$~\cite{lee2011quantification} are evaluated to measure nonclassicality and quantum coherence, respectively, which are defined as
\begin{align}
	\delta&=\int[|W(\alpha,\alpha^*)|-W(\alpha,\alpha^*)]d^2\alpha>0,\nonumber\\
	I&=\frac{\pi}{2}\int W(\alpha,\alpha^*)( -\frac{\partial^2}{\partial\alpha\partial\alpha^*}-1 )W(\alpha,\alpha^*)d^2\alpha>0.
	\label{quantumness}
\end{align}
Larger values of $\delta$ and $I$ indicate a better quality of quantum cat, and zeroes of them imply vanishing of cat state no matter how large $|\alpha|^2$ gets.

Figure~\ref{fig:F-r} shows the effect of the magnetic damping $\gamma$ on the above properties, for the generated odd (a) and even (b) cat states, at a fixed value of effective squeezing parameter $r=0.2$. It is obvious that the nonclassical properties of $F$, $\delta$ and $I$ decrease drastically as damping $\gamma$ increases. This is not surprising since a higher damping $\gamma$ corresponds to a larger decoherence. For $g/2\pi=5\,$MHz, when the decay $\gamma/2\pi>0.16\,$MHz, $\delta$ and $I$ reach $0$ where coherence in the superposition disappears. As $\gamma$ increases, the value of $G=g^2/\kappa-\gamma$ decreases. Hence the fixed $r=G\tau$ requires longer interaction time $\tau$, and leads to more magnon excitations with larger $|\alpha|^2$. This correspond to a classical mixture of two components resulting from interaction with environment, revealing the quantum-to-classical transition. Consequently, one need to choose appropriate parameters carefully to ensure both nonclassical properties of $F$, $\delta$, $I$ and the size of cat $|\alpha|^2$ as large as possible. Clearly, as the above quantities depend on the effective squeezing parameter $r=G\tau$ with $G=g^2/\kappa-\gamma$, we have that the tolerance to damping $\gamma$ can be improved by enhancing the optomagnonic coupling $g$~\cite{supplementary}. 

The gray vertical lines in Fig.~\ref{fig:F-r} correspond to the parameters studied in Fig.~\ref{fig:cats_in_time} (b), where an odd (even) magnon cat state of size $|\alpha|^2=1.44~(3.42)$ is remotely created. The size corresponds to an overlap between the two peaks of $5.6\%$ ($0.1\%$) for the odd (even) cat, which is adequate for exploiting such states in practical applications, e.g.~fault-tolerant quantum computing~\cite{lund2008fault}. At the same time, high fidelity for the odd (even) cat $F=0.85~(0.70)$, Wigner negativity $\delta=0.30~(0.40)$, and macroscopic quantum superposition $I=0.82~(0.91)$ are also achieved.  Besides, both the imperfections of projective measurement $|X_C|\le\epsilon=0.1$ and the dark counts when performing single-photon operations are considered, where magnon cat states still maintain well with a negligible reduction in above properties (Figs. S4 and S5~\cite{supplementary}).  Moreover, the lifetime under these practical considerations is expected to be $\sim1.12\,(0.54)\,\mu$s for the odd (even) magnon cat state (Fig.~S6~\cite{supplementary}).

\textit{Measurement of magnon cat states.--} 
     Lastly, we discuss how to detect the created magnon cat state. Since magnon modes in YIG can coherently couple to optical and microwave modes, the read-out of their states can be implemented through these two channels. The optical method can be realized in analogy to optomechanics~\cite{palomaki2013coherent}, where the magnon cat state can be mapped onto the cavity field pumped by a red-detuned tone with $\Delta=\omega_m$ (resulting in a beamsplitter-type interaction)~\cite{bittencourt2019magnon}, and then performing quantum tomography on the output optical mode. The microwave method is to place the YIG sphere inside a microwave resonant cavity, and measure in the strong coupling regime the output microwave signal which carries information on the magnon state~\cite{braggio2017optical,*braggio2021direct,zhang2017observation,grigoryan2018synchronized}. Alternatively, the magnon can be coupled via microwave photons to a superconducting qubit, which is then read-out spectroscopically as demonstrated for single magnon detection~\cite{tabuchi2015coherent,lachance2017resolving,lachance2020entanglement}.

         In summary, we have proposed a scheme to remotely generate and manipulate magnon Schr\"{o}dinger cat states via EPR steering in a pulsed optomagnonic system. The magnon state in a YIG sphere at one position collapses into an odd (even) cat state by applying single-photon subtraction (followed by a single-photon addition) on the distant optical mode that previously entangled with it. This remote creation method is distinct from the conventional approach that prepares nonclassical states by local operations on the magnon mode, and thus paves the way for future magnon-based quantum internet and quantum computing research. It is also promising to facilitate the quantum precision measurements, such as dark-matter searches~\cite{flower2019broadening,*marsh2019proposal}. By examining key properties of the created magnon cat states, we find that the cat is of relatively large size with high fidelity and quantum coherence, and can survive $t\sim1\,\mu$s, even in the presence of dissipation resulting from the decay of modes, the error of projective measurements, and the dark counts when performing single-photon operations.  
         
         Note that recently, significant efforts have been employed to improve the optomagnonic coupling by enhancing the spatial overlap of whispering gallery and surface magnetostatic modes~\cite{sharma2019optimal}, or by reducing the modal volume with a YIG microdisk~\cite{graf2018cavity}. Suppose that the optomagnonic coupling can be enhanced to $g=15\,$MHz, the requirement on the decay of magnon mode for creating cat states can be relaxed by about an order of magnitude $~\gamma/2\pi<1.6\,$MHz (Fig.~S2~\cite{supplementary}). Besides ferromagnetic systems studied in this work, the paramagnetic systems with lower damping and weaker coupling are also candidates for our scheme, if $G=g^2/\kappa-\gamma>0$ can be achieved in experiment. The approach can be also applied to other physical platforms, such as linear optical network, optomechanical systems, and other hybrid quantum systems.

\begin{acknowledgments}
	We thank the constructive discussions with M. Fadel. This work is supported by the National Natural Science Foundation of China (Grants No.~61675007, No. 11975026, and No.~61974067),  the Key R\&D Program of Guangzhou Province (Grant No.~2018B030329001), and the Beijing Natural Science Foundation (Grant No. Z190005). F.-X. S. acknowledges the China Postdoctoral Science Foundation (Grant No.~2020M680186). K.X. is supported by the NSFC (Grants No.~61774017, No.~11734004, and No.~12088101), and NSAF (Grant No.~U1930402).
\end{acknowledgments}

\bibliography{cat-magnon}

\end{document}


\title{Supplemental Material on ``Remote generation of magnon Schr\"{o}dinger cat state via magnon-photon entanglement''}

\author{Feng-Xiao Sun}
\affiliation{State Key Laboratory for Mesoscopic Physics, School of Physics, Frontiers Science Center for Nano-optoelectronics, $\&$ Collaborative
Innovation Center of Quantum Matter, Peking University, Beijing 100871, China}
\affiliation{Collaborative Innovation Center of Extreme Optics, Shanxi University, Taiyuan, Shanxi 030006, China}
\author{Sha-Sha Zheng}
\affiliation{State Key Laboratory for Mesoscopic Physics, School of Physics, Frontiers Science Center for Nano-optoelectronics, $\&$ Collaborative
Innovation Center of Quantum Matter, Peking University, Beijing 100871, China}
\affiliation{Collaborative Innovation Center of Extreme Optics, Shanxi University, Taiyuan, Shanxi 030006, China}
\author{Yang Xiao}
\affiliation{Department of Applied Physics, Nanjing University of Aeronautics and Astronautics, Nanjing 210016, China}
\author{Qihuang Gong}
\affiliation{State Key Laboratory for Mesoscopic Physics, School of Physics, Frontiers Science Center for Nano-optoelectronics, $\&$ Collaborative
Innovation Center of Quantum Matter, Peking University, Beijing 100871, China}
\affiliation{Collaborative Innovation Center of Extreme Optics, Shanxi University, Taiyuan, Shanxi 030006, China}
\affiliation{Peking University Yangtze Delta Institute of Optoelectronics, Nantong 226010, Jiangsu, China}
\author{Qiongyi He}
\email{qiongyihe@pku.edu.cn}
\affiliation{State Key Laboratory for Mesoscopic Physics, School of Physics, Frontiers Science Center for Nano-optoelectronics, $\&$ Collaborative
Innovation Center of Quantum Matter, Peking University, Beijing 100871, China}
\affiliation{Collaborative Innovation Center of Extreme Optics, Shanxi University, Taiyuan, Shanxi 030006, China}
\affiliation{Peking University Yangtze Delta Institute of Optoelectronics, Nantong 226010, Jiangsu, China}
\author{Ke Xia}
\email{kexia@csrc.ac.cn}
\affiliation{Beijing Computational Science Research Center, Beijing 100193, China}

\maketitle

\section{Optomagnonic coupling}
The couplings between the optical mode and the magnon mode have been experimentally realized in the optomagnonic systems~\cite{osada2016cavity,zhang2016optomagnonic,haigh2016triple}, where the Hamiltonian reads~\cite{kusminskiy2016coupled},
\begin{equation}
	H=\hbar\Delta c^{\dagger}c-\hbar\omega_{m} {S}_{z}+\hbar G_m {S}_{x}c^{\dagger}c.
	\label{Hamiltonian_spin}
\end{equation}
Here, $c$ is the annihilation operator for the optical mode, $\Delta=\omega_{c}-\omega_{p}$ is the detuning of the cavity with respect to the pumped laser frequency $\omega_{p}$, $\textbf{S}=({S}_{x},{S}_{y},{S}_{z})$ is the macrospin operator, with magnetization axis along $z$ and precession frequency $\omega_{m}$ that can be controlled by an external magnetic field. The homogeneous magnon mode couples to the optical mode with a parametric optomagnonic coupling coefficient $G_m$ which is a material-dependent constant. For a polarized state, we can adopt the Holstein-Primakoff transformation of the spin operators~\cite{holstein1940field}, 
\begin{align}
	S_{+}&=\sqrt{2S-m^{\dagger}m}m,\nonumber\\
	S_{-}&=m^{\dagger}\sqrt{2S-m^{\dagger}m},\nonumber\\
	S_{z}&=S-m^{\dagger}m,
	\label{HP}
\end{align}
where $S$ is the total spin number of the YIG sphere, and $S_{+}=S_{x}+iS_{y},~S_{-}=S_{x}-iS_{y}$ are the raising and lowering operators of the
macrospin. The mean number of spin excitations is expected to be much smaller than the total spin, i.e.~$\langle mm^\dagger\rangle \ll S$, and therefore the spin operators can be safely approximated to bosonic annihilation and creation operators $S_+\approx\sqrt{2S}m$, $S_-\approx\sqrt{2S}m^\dagger$, with the commutation relation for boson operators $[m,m^\dagger]=1$. Then the Hamiltonian (\ref{Hamiltonian_spin}) can be expressed as
\begin{equation}
	H=\hbar\Delta c^{\dagger}c+\hbar\omega_{m}m^{\dagger}m+\hbar G_m\sqrt{S/2}c^{\dagger}c(m+m^{\dagger}).
	\label{Hamiltonian_linear}
\end{equation}
This becomes formally equivalent to the optomechanical interaction, as given by Hamiltonian (1) in the main text with single-photon coupling constant $g_{0}=G_m\sqrt{S/2}$.

We pump the cavity mode by a blue-detuned laser with $\Delta=-\omega_{m}$ and  write the mode operator as composed of a large classical amplitude and a small fluctuation operator, i.e. $c \rightarrow \beta + c$. The linearized Hamiltonian is given as
\begin{equation}
	H=-\hbar\omega_{m} c^{\dagger}c+\hbar\omega_{m}m^{\dagger}m+\hbar g_{0}|\beta|(mc+m^{\dagger}c^{\dagger}+mc^{\dagger}+m^{\dagger}c).
\end{equation}
In a frame rotating with $\omega_m$, $m'=me^{i\omega_{m}t}$, $c'=ce^{-i\omega_{m}t}$, and under the rotating-wave approximation in which we ignore all terms oscillating with $2\omega_m$, the Hamiltonian is dominated by the parametric-down-conversion-type (two-mode squeezing) interaction,
\begin{equation}
	H_{\text{RWA}}=\hbar g (m'c'+{m'}^{\dagger}{c'}^{\dagger}),
\end{equation}
where $g=g_{0}|\beta|$ is the effective optomagnonic coupling constant. For clarity of the notation, we have replaced $m'\to m$ and $c'\to c$ in the Hamiltonian (2) in the main text.

As there are always inevitable decays for both the optical and the magnon modes, we study the evolution of the system by linearized quantum Langevin equations,
\begin{align}
	\dot{m}&=-\gamma m-ig c^{\dagger}-\sqrt{2\gamma}m_{in},\nonumber\\
	\dot{c}&=-\kappa c-ig m^{\dagger}-\sqrt{2\kappa}c_{in}.
	\label{EOM}
\end{align}
The dynamics of operators is affected by input noises $m_{in},~c_{in}$
arising from the coupling of the modes to their surrounding
environments, characterized by the correlation functions $\langle m_{in}(t)m_{in}^{\dagger}(t') \rangle=(n_m+1)\delta(t-t')$ and $\langle c_{in}(t)c_{in}^{\dagger}(t') \rangle=(n_c+1)\delta(t-t')$. Here we are considering that the system is in the low temperature $T$ so that $\hbar\omega_{c}\gg\hbar\omega_{m}\gg k_BT$, where $k_B$ is the Boltzmann constant. Hence, the average number of excitations due to thermal noise in the environment coupled to the mode is $n_{c(m)}=1/[\exp(\hbar\omega_{c(m)}/k_BT)-1]\approx0$. Specifically, for the considered magnon mode $\omega/2\pi=10$ GHz, it will be cooled to near the ground state at a low temperature $T<0.7\,$K, where a negligible thermal occupation $n_{m}<0.1$ corresponds to $T<0.2\,$K. Such temperature is accessible by using a dilution refrigerator with low magnon damping~\cite{tabuchi2014hybridizing}. And the effects of the thermal occupation are discussed in SIII.

We make an adiabatic approximation, $\dot{c}=0$, which is justified for the condition $\kappa\gg\gamma,~g$, satisfying the implementation of experiments~\cite{osada2016cavity,zhang2016optomagnonic,haigh2016triple}. Thus the optical mode can be solved analytically, namely
\begin{equation}
	c=-i\frac{g}{\kappa}m^{\dagger}-\sqrt{\frac{2}{\kappa}}c_{in},
\end{equation}
and the magnon's equation of motion is simplified as,
\begin{equation}
	\dot{m}=Gm+ig\sqrt{\frac{2}{\kappa}}c_{in}^{\dagger}-\sqrt{2\gamma}m_{in},
\end{equation}
where $G=g^2/\kappa-\gamma$. Since the optical mode is pumped by pulse in our proposal, it is convenient to introduce normalized temporal modes of the input and output cavity fields, given by the standard input-output relation, $c_{out}(t)=c_{in}(t)+\sqrt{2\kappa}c(t)$. The normalized temporal operators for the optical field are defined by~\cite{vanner2011pulsed,hofer2011quantum} 
\begin{align}
	C_{in}&=\sqrt{\frac{2G}{1-e^{-2G\tau}}}\int_0^{\tau}e^{-Gt}c_{in}(t)dt,\nonumber\\
	\tilde{C}_{in}&=\sqrt{\frac{2G}{e^{2G\tau}-1}}\int_{0}^{\tau}e^{Gt}c_{in}(t)dt,\nonumber\\
	C_{out}&=\sqrt{\frac{2G}{e^{2G\tau}-1}}\int_{0}^{\tau}e^{Gt}c_{out}(t)dt,
	\label{temporal_mode_c}
\end{align}
where $\tau$ is the duration of the laser pulse, which leads to the effective squeezing parameter together with coupling $G$, $r=G\tau$. We may also define the normalized operators for the magnon mode, 
\begin{align}
	M_m&=\sqrt{\frac{2G}{1-e^{-2G\tau}}}\int_{0}^{\tau}e^{-Gt}m_{in}(t)dt,\nonumber\\
	\tilde{M}_m&=\sqrt{\frac{2G}{e^{2G\tau}-1}}\int_{0}^{\tau}e^{Gt}m_{in}(t)dt,\nonumber\\
	M_{in}&=m(0),\;M_{out}=m(\tau).
	\label{temporal_mode_m}
\end{align}
At the end of the pulse, $t=\tau$, the optical and the magnon modes arrive at,
\begin{align}
	C_{out}&=-i\sqrt{\frac{2g^2}{\kappa}}\sqrt{\frac{e^{2r}-1}{2G}}M_{in}^{\dagger}-\tilde{C}_{in}-\frac{g^2}{G\kappa}(C_{in}e^r-\tilde{C}_{in})+i\frac{g}{G}\sqrt{\frac{\gamma}{\kappa}}(M_m^\dagger e^r-\tilde{M}_m^\dagger),\nonumber\\
	M_{out}&=e^{r}M_{in}+\sqrt{\frac{e^{2r}-1}{2G}}(i\sqrt{\frac{2g^2}{\kappa}}C_{in}^\dagger-\sqrt{2\gamma}M_m).
\end{align}
The solution for the output modes $X_{C,M}$ and $P_{C,M}$ can be derived directly, which are used to analyze the following Wigner function of the cat state as well as magnon-photon entanglement and EPR steering.

\section{Single-photon operations}
In our scheme, single-photon operations are applied on the remote output optical field which can steer the magnon mode. Experimentally, the single-photon subtraction can be realized by a beam splitter with high transmission and low reflection such that a single-photon is probabilistically subtracted from the optical mode~\cite{wenger2004non,parigi2007probing,ra2020non}. A click on the photodetector in the reflected path indicates successful single-photon subtraction from the optical mode, which leads to the density matrix $\rho_{sub}\propto a\rho a^{\dagger}$. 

In the following we derive the Wigner functions of the state with single-photon subtraction. The Wigner function represents a typical phase-space quasiprobability distribution, which gives us the joint probability distribution of the quadratures for the quantum state~\cite{wigner1932on}. Given the density matrix of the quantum state, $\rho$, the corresponding Wigner function is defined as,
\begin{equation}
	W(X, Y)=\frac{1}{\pi \hbar} \int e^{-2 i x Y / \hbar}\langle X-x|\rho| X+x\rangle d x.
\end{equation}

Expressing the quadratures by the coherent amplitudes,
\begin{equation}
	X=\frac{\alpha+\alpha^{*}}{\sqrt{2}},\; Y=\frac{\alpha-\alpha^{*}}{\sqrt{2} i}, 
\end{equation}
we can rewrite the Wigner function as the Fourier transform of the characteristic function $C(\beta,\beta^{*})$,
\begin{equation}
	W\left(\alpha, \alpha^{*}\right)=\frac{1}{\pi^{2}} \int C\left(\beta, \beta^{*}\right) e^{-i \beta^{*} \alpha-i \beta \alpha^{*}} d^{2} \beta,
	\label{wigner_fourier}
\end{equation}
where
\begin{equation}
	C\left(\beta, \beta^{*}\right)=\operatorname{Tr}\left(e^{i \beta^{*} a+i \beta a^{\dagger}} \rho\right)=\int c_{\xi,\xi'}\langle \xi|e^{i\beta^*a+i\beta a^{\dagger}}|\xi' \rangle d^2\xi,
	\label{characteristic}
\end{equation}
with $c_{\xi,\xi'}=\langle \xi'|\rho|\xi \rangle$. By using Glauber equation~\cite{scully2012quantum}, $e^{A+B}=e^{-[A,B]/2}e^{A}e^{B}$, the characteristic function can be further simplified as,
\begin{equation}
	C(\beta,\beta^{*})=\int c_{\xi,\xi'} e^{-\frac{|\beta|^2}{2}}e^{i\beta\xi^*}e^{i\beta^{*}\xi'}\langle \xi|\xi' \rangle d^2\xi.
\end{equation}
Here we have used $a|\xi\rangle=\xi|\xi\rangle$.

For the state after single-photon subtraction, we have 
\begin{equation}
	C_{sub}(\beta,\beta^{*})=\int c_{\xi,\xi'}\langle \xi|a^{\dagger}e^{i\beta^*a+i\beta a^{\dagger}}a|\xi' \rangle d^2\xi=\int e^{-\frac{|\beta|^2}{2}}e^{i\beta\xi^*}e^{i\beta^{*}\xi'}\xi^*\xi'\langle \xi|\xi' \rangle d^2\xi.
\end{equation}
Considering that
\begin{align}
	\frac{\partial}{\partial\beta}\langle \xi|e^{i\beta^*a+i\beta a^{\dagger}}|\xi' \rangle=\left( -\frac{\beta^*}{2}+i\xi^* \right)\langle \xi|e^{i\beta^*a+i\beta a^{\dagger}}|\xi' \rangle,\nonumber\\
	\frac{\partial}{\partial\beta^*}\langle \xi|e^{i\beta^*a+i\beta a^{\dagger}}|\xi' \rangle=\left( -\frac{\beta}{2}+i\xi' \right)\langle \xi|e^{i\beta^*a+i\beta a^{\dagger}}|\xi' \rangle,
\end{align}
or equally,
\begin{align}
	i\xi^*\langle \xi|e^{i\beta^*a+i\beta a^{\dagger}}|\xi' \rangle=\left( \frac{\partial}{\partial\beta}+\frac{\beta^*}{2} \right)\langle \xi|e^{i\beta^*a+i\beta a^{\dagger}}|\xi' \rangle=\left( \frac{\partial}{\partial\beta}+\frac{\beta^*}{2} \right)e^{-\frac{|\beta|^2}{2}}e^{i\beta\xi^*}e^{i\beta^{*}\xi'}\langle \xi|\xi' \rangle,\nonumber\\
	i\xi'\langle \xi|e^{i\beta^*a+i\beta a^{\dagger}}|\xi' \rangle=\left( \frac{\partial}{\partial\beta^*}+\frac{\beta}{2} \right)\langle \xi|e^{i\beta^*a+i\beta a^{\dagger}}|\xi' \rangle=\left( \frac{\partial}{\partial\beta^*}+\frac{\beta}{2} \right)e^{-\frac{|\beta|^2}{2}}e^{i\beta\xi^*}e^{i\beta^{*}\xi'}\langle \xi|\xi' \rangle,
\end{align}
we find relation between the characteristic functions before and after the single-photon subtraction,
\begin{equation}
	C_{sub}(\beta,\beta^{*})=-\left(\frac{\partial}{\partial\beta}+\frac{\beta^*}{2}\right)\left( \frac{\partial}{\partial\beta^*}+\frac{\beta}{2} \right)C(\beta,\beta^*).
	\label{characteristic_sub}
\end{equation}
Therefore, the Wigner functions after the single-photon subtraction are directly obtained via the Fourier transform~(\ref{wigner_fourier}).

As for the single-photon addition, it can be experimentally realized by the conditional stimulated parametric down-conversion in a nonlinear optical crystal~\cite{zavatta2004quantum,parigi2007probing}.
The spontaneous parametric down-conversion takes place with the nonlinear optical crystal, where a high-frequency pump photon will transfer to two low-frequency photons, namely the signal and the idler modes. By injecting the output optical mode of our proposal into the the signal mode, the single-photon addition will be applied every time that a single photon is detected in the idler mode. After the single-photon addition, the density matrix reads $\rho_{add}\propto a^{\dagger}\rho a$.
With the similar method, we will obtain the relation between the characteristic functions before and after the single-photon addition,
\begin{equation}
	C_{add}(\beta,\beta^{*})=-\left(\frac{\partial}{\partial\beta}-\frac{\beta^*}{2}\right)\left( \frac{\partial}{\partial\beta^*}-\frac{\beta}{2} \right)C(\beta,\beta^*).
	\label{characteristic_add}
\end{equation}
And the Wigner functions can be derived by using Fourier transform~(\ref{wigner_fourier}).

\section{Transient magnon cat states}
           In the Fig.~3 in the main text, we have shown the Wigner functions of the transient magnon cat states at some time slices. The full time evolution of the transient magnon cat states can be found in the video attached here, named as ``Supplement\_cat\_time.avi''. The left column shows the odd magnon cat state with a single-photon subtraction applied on the distant optical field, while the right column shows the even magnon cat state achieved by a single-photon subtraction and followed by a single-photon addition applied. The figures in the upper row are the $3$D Wigner functions of the magnon cat states, and those in the lower row are the corresponding $2$D projection. From the video, we find that both the odd and the even cat states grow up from the vacuum states, and the interference fringes become less apparent with long duration of the optical pulse.
 \\
           
The quality of the transient magnon cat states changing with effective squeezing parameter $r=G\tau$ is indicated in Fig.~\ref{fig:F-t}, where we investigate the following figures of merit: the cat size $|\alpha|^2$, fidelity $F$ between the produced state and an ideal Schr\"{o}dinger cat, Wigner negativity $\delta$, and macroscopic quantum superposition $I$, defined in the main text. It is obvious that as damping $\gamma$ increases, the cat size increases, while the nonclassical properties of $F$, $\delta$ and $I$ sharply increase within short time evolution and then eventually decrease with long time of interaction due to the environment-induced decoherence. Consequently, it is important to choose proper optical pulse duration $\tau$ to ensure both nonclassical properties of $F$, $\delta$, $I$ and the size of cat $|\alpha|^2$ as large as possible for a good quality of quantum cat. 
\begin{figure}
	\centering
	\includegraphics[width=0.238\textwidth]{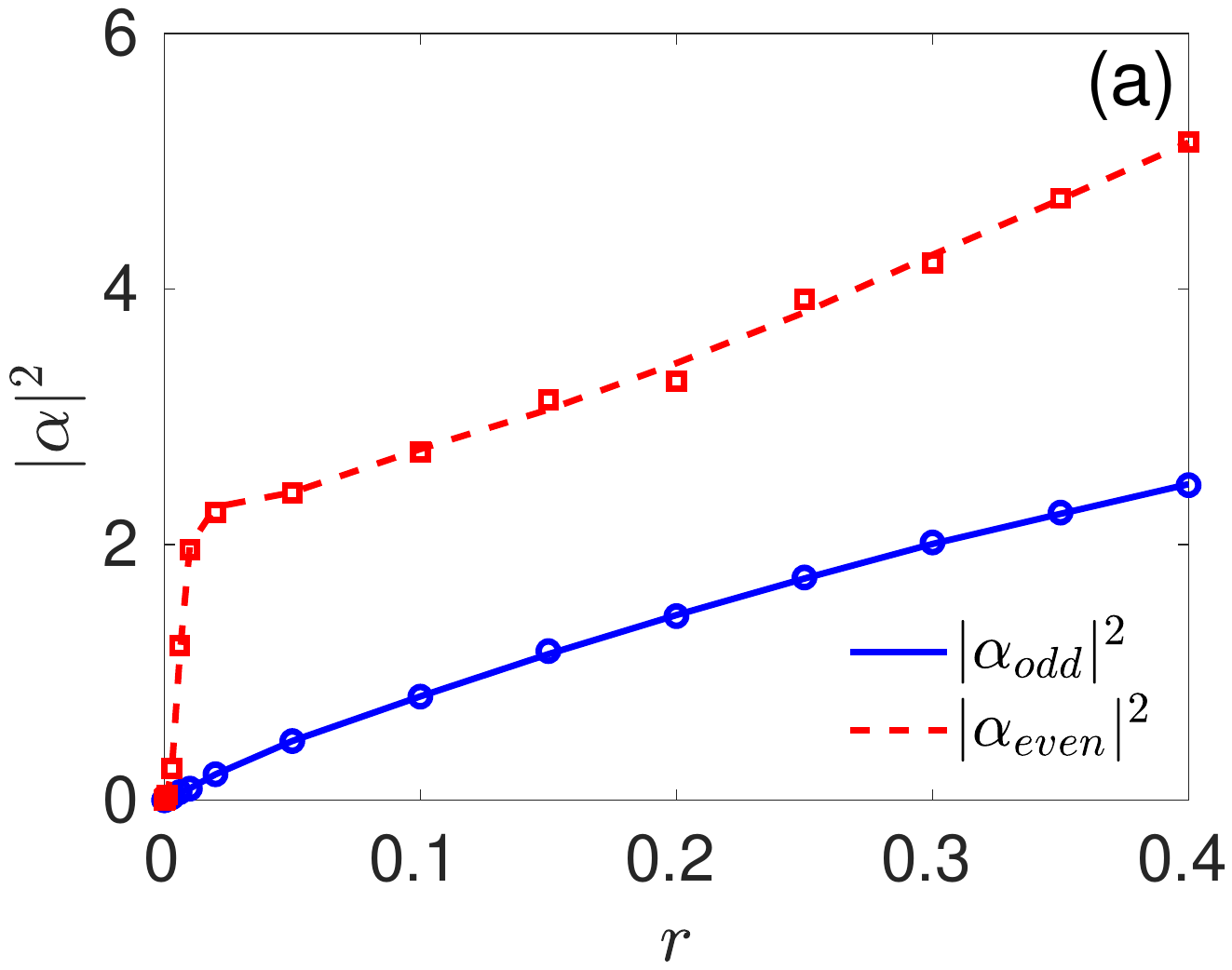}
	\includegraphics[width=0.245\textwidth]{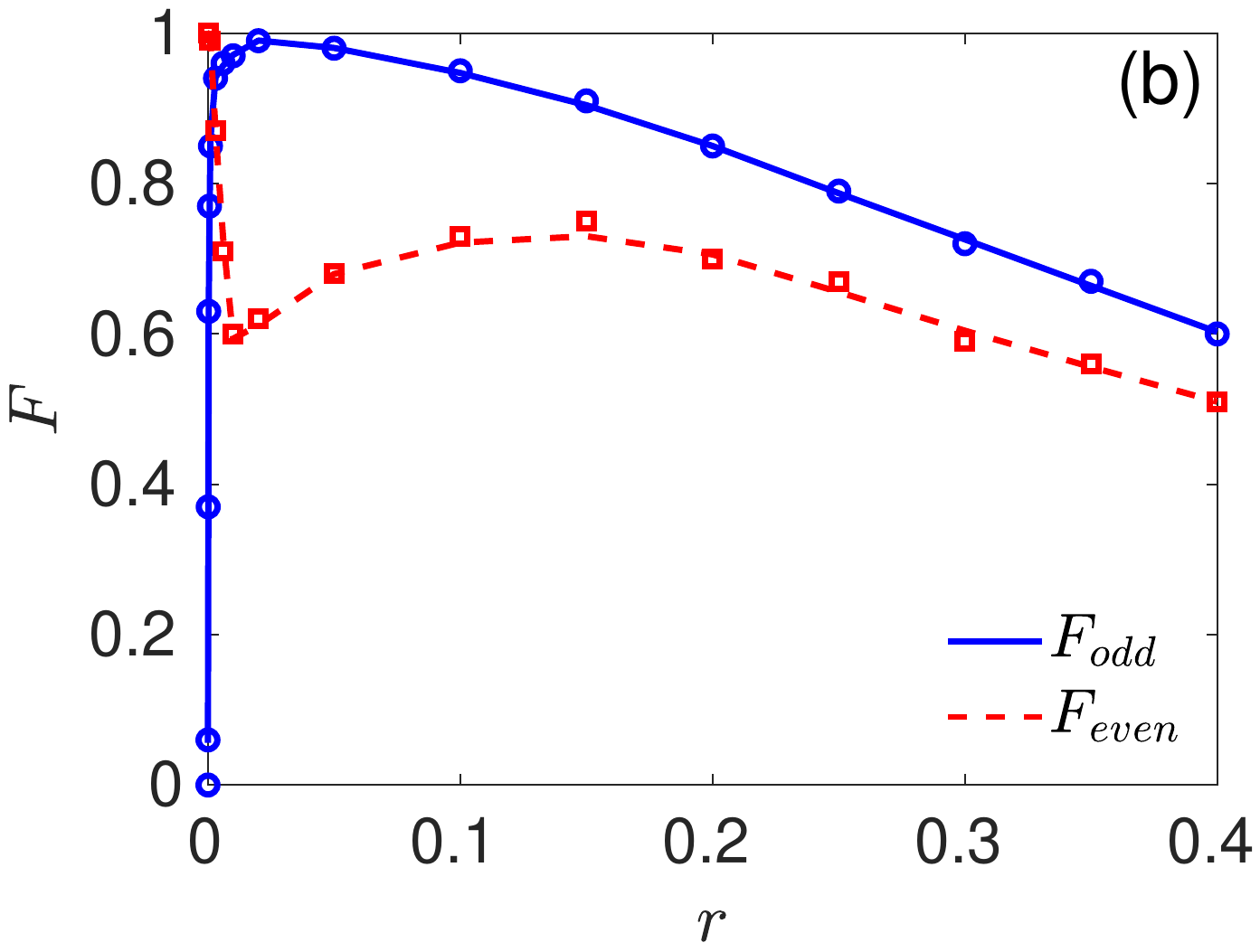}
	\includegraphics[width=0.245\textwidth]{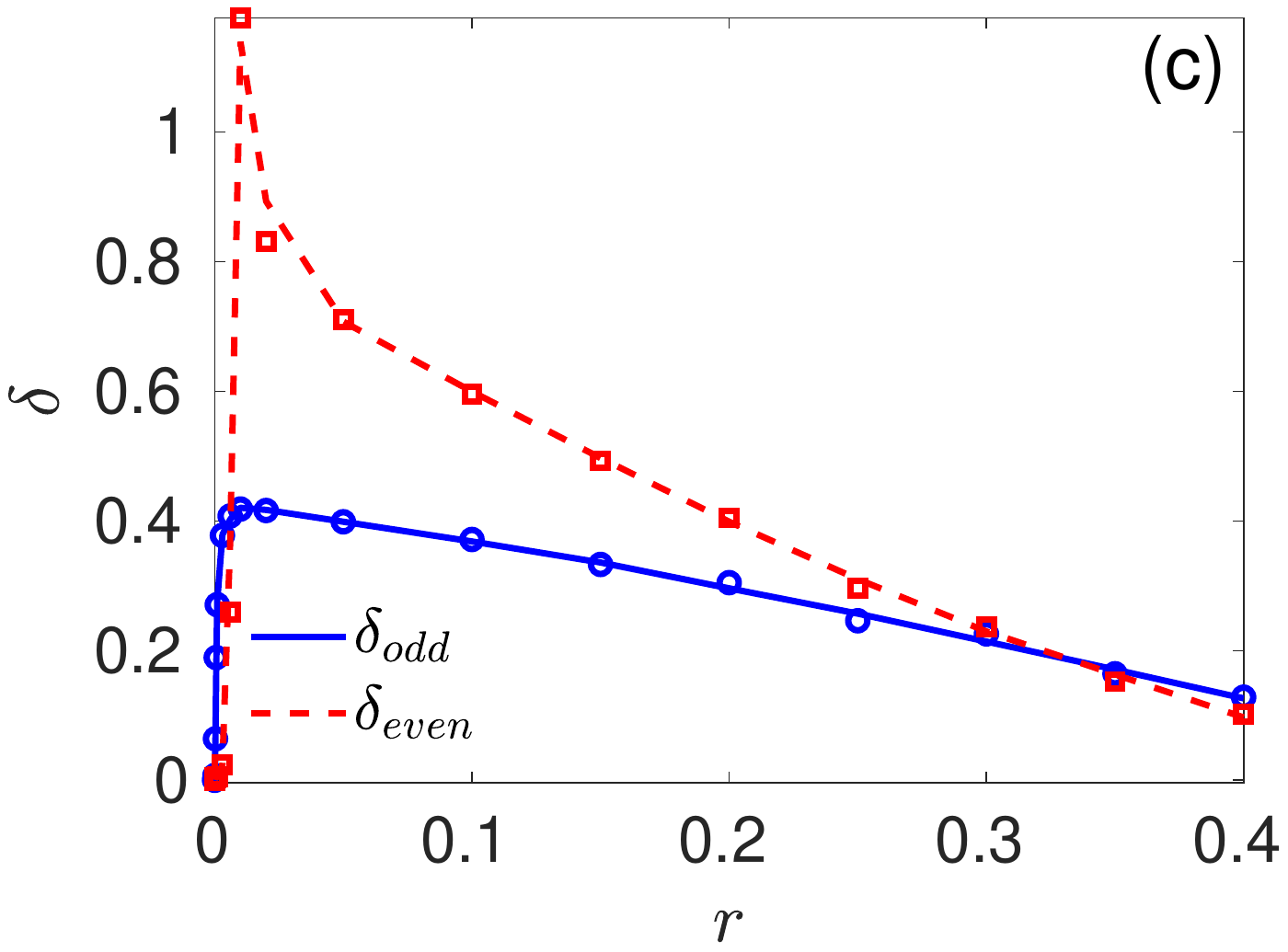}
	\includegraphics[width=0.245\textwidth]{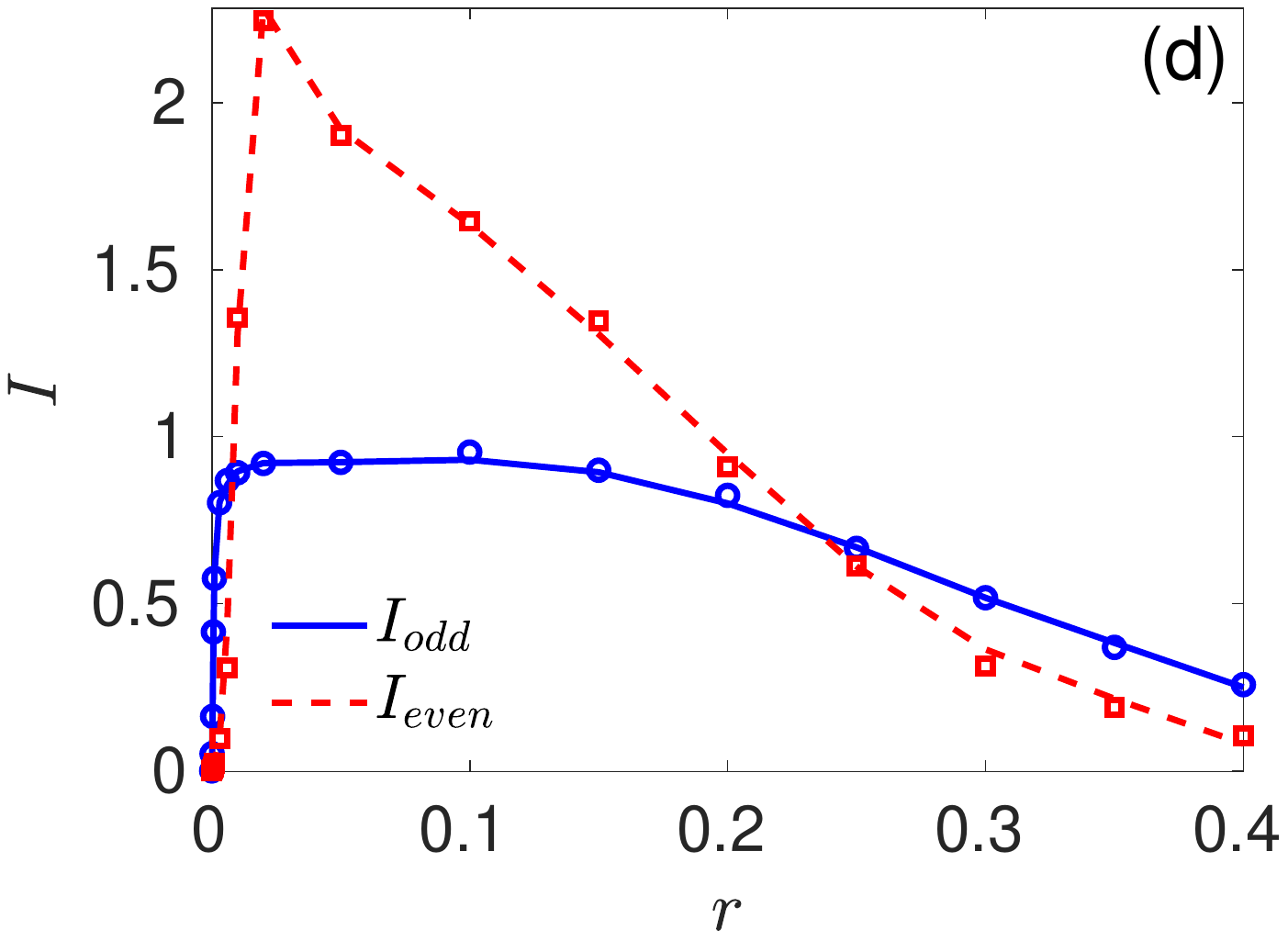}
	\caption{(a) Cat `size' $|\alpha|^2$, (b) fidelity $F$, (c) Wigner negativity $\delta$, and (d) the macroscopic quantum superposition $I$ of the state remotely created through EPR steering by single-photon operation and a projective measurement for $X_C=0$ performed on the optical field. The blue solid and red dashed fitted curves correspond to the magnon odd and even cat states, respectively. The parameters are the same as Fig.~3 in the main text.}
	\label{fig:F-t}
\end{figure}
\\

\begin{figure}
	\centering
	\includegraphics[width=0.35\textwidth]{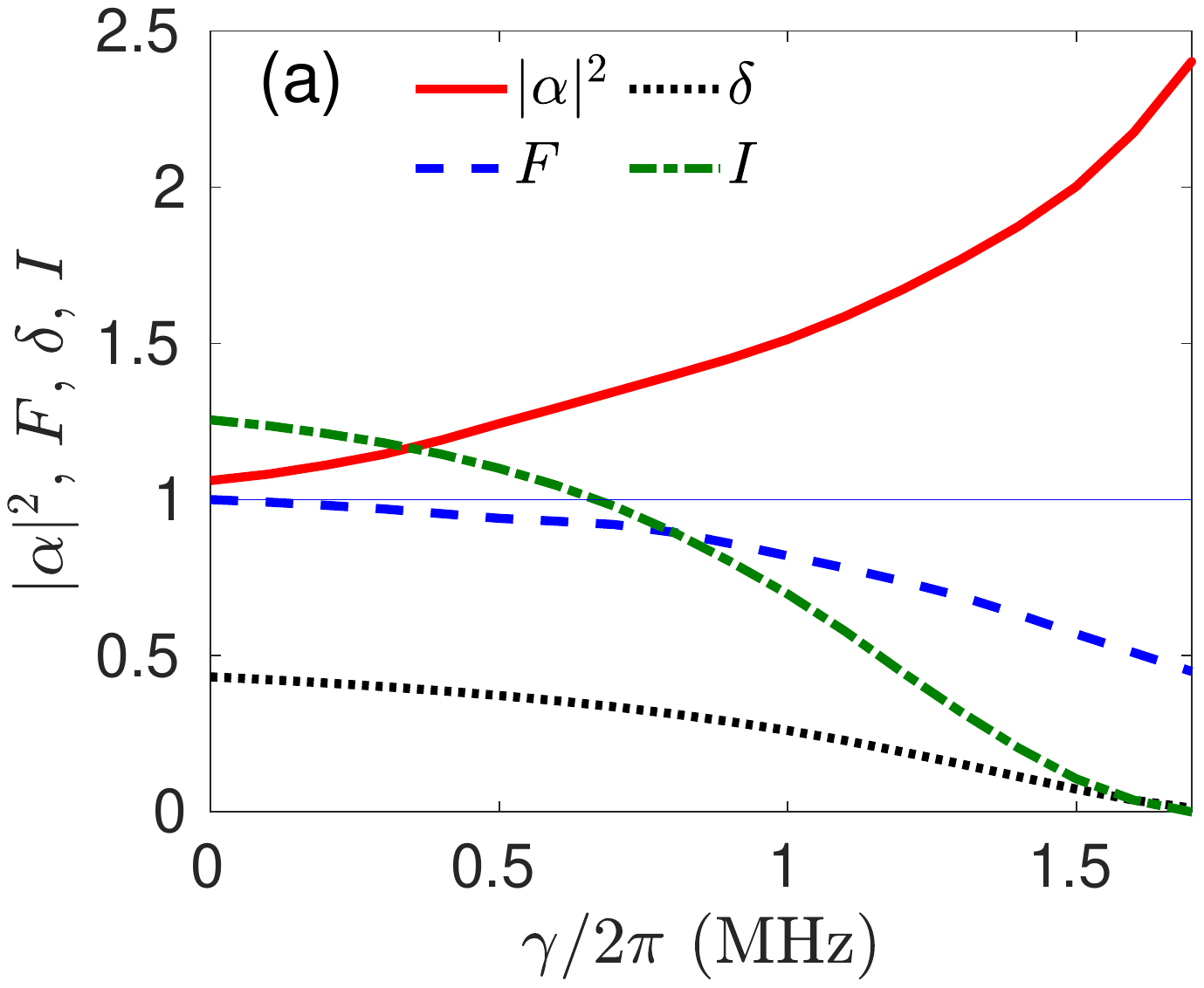}
	\includegraphics[width=0.34\textwidth]{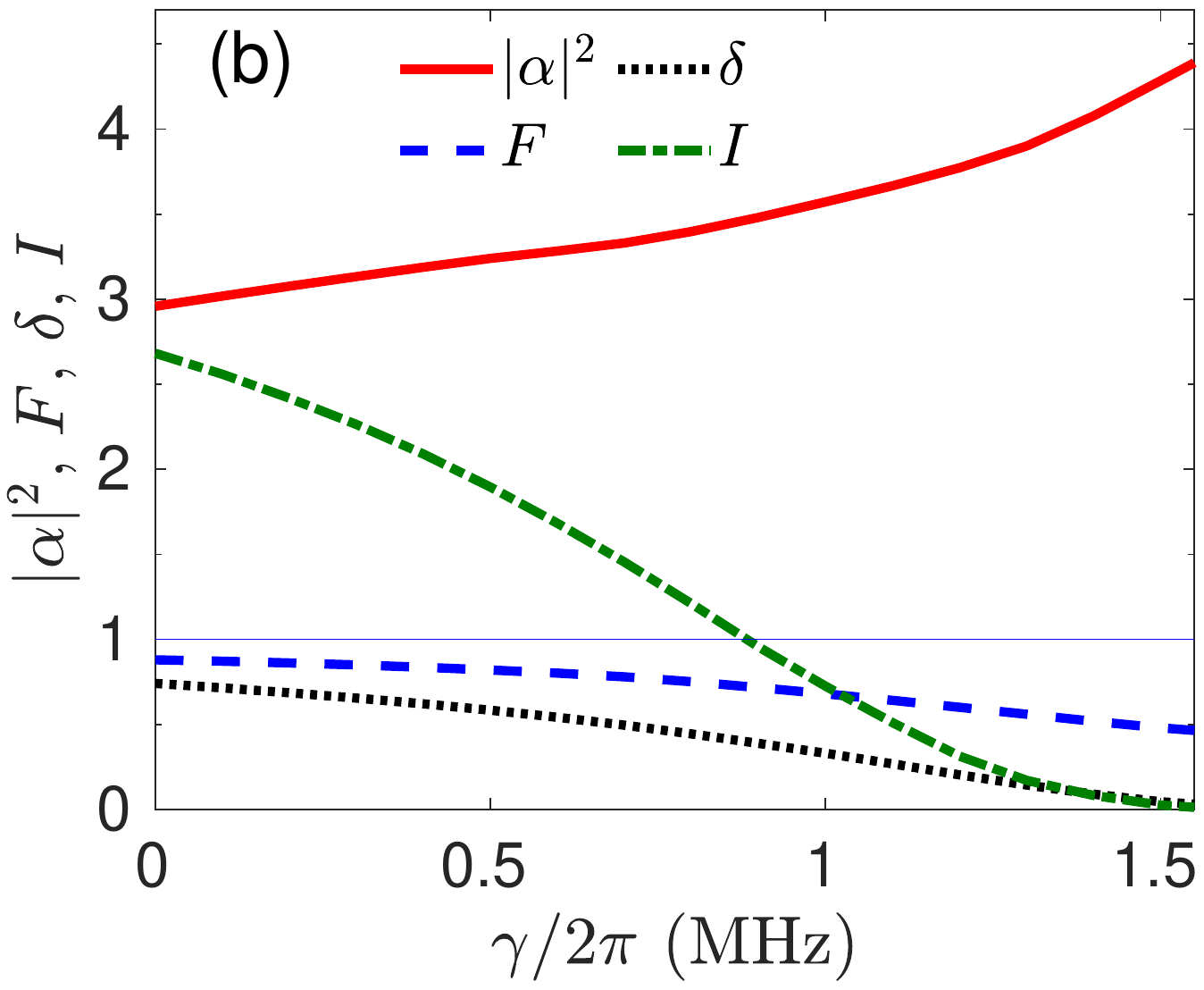}
	\caption{Cat size $|\alpha|^2$ (red solid), fidelity $F$ (blue dashed), Wigner negativity $\delta$ (black dots), and macroscopic quantum superposition $I$ (green dash-dotted) of (a) the odd cat states and (b) the even cat states remotely created varying with $\gamma$ at $r=G\tau=0.2$ with $g/2\pi=15\,$MHz and $\kappa/2\pi=100\,$MHz.}
	\label{fig:F-gamma}
\end{figure}
Besides, the above quantities clearly depend on the effective coupling  $G=g^2/\kappa-\gamma$. Figure 4 in the main text has shown the effect of the magnetic damping $\gamma$ on the above properties for a fixed value of $g/2\pi=5\,$MHz. The results shown there can be further improved by enhancing the optomagnonic coupling $g$ to offset the decoherence introduced by $\gamma$. For instance, if the optomagnonic coupling can be enhanced to $g/2\pi=15\,$MHz by improving the quality of the optical cavity and the YIG, the requirement on the decay can be relaxed by about an order of magnitude from $\gamma/2\pi<0.16\,$MHz to $~\gamma/2\pi<1.6\,$MHz, as shown in Fig.~\ref{fig:F-gamma}. By choosing proper optical pulse  $r=0.2$ with $g/2\pi=15\,$MHz, $\gamma=1\,$MHz and $\kappa=100\,$MHz, a relative large odd (even) magnon cat state of size $|\alpha|^2=1.56~(3.61)$, fidelity $F=0.82~(0.68)$, Wigner negativity $\delta=0.26~(0.33)$, macroscopic quantum superposition $I=0.70~(0.72)$ is remotely created. This corresponds to an overlap between the two peaks of $4.42\%$ and $0.07\%$ for the odd and even cat states, respectively, which are adequate for exploiting such states in practical applications.
\\

\begin{figure}
	\centering
	\includegraphics[width=0.35\textwidth]{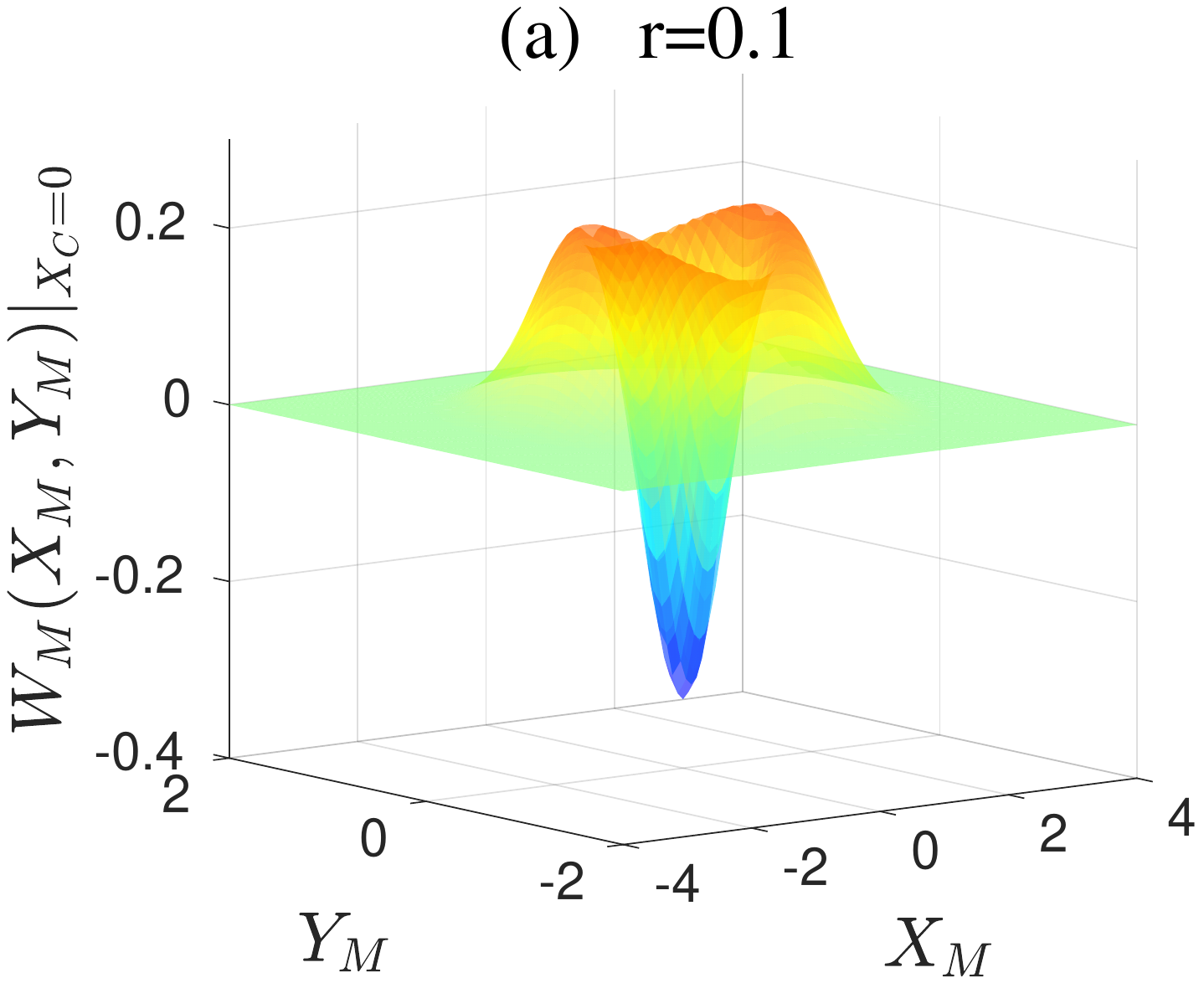}
	\includegraphics[width=0.35\textwidth]{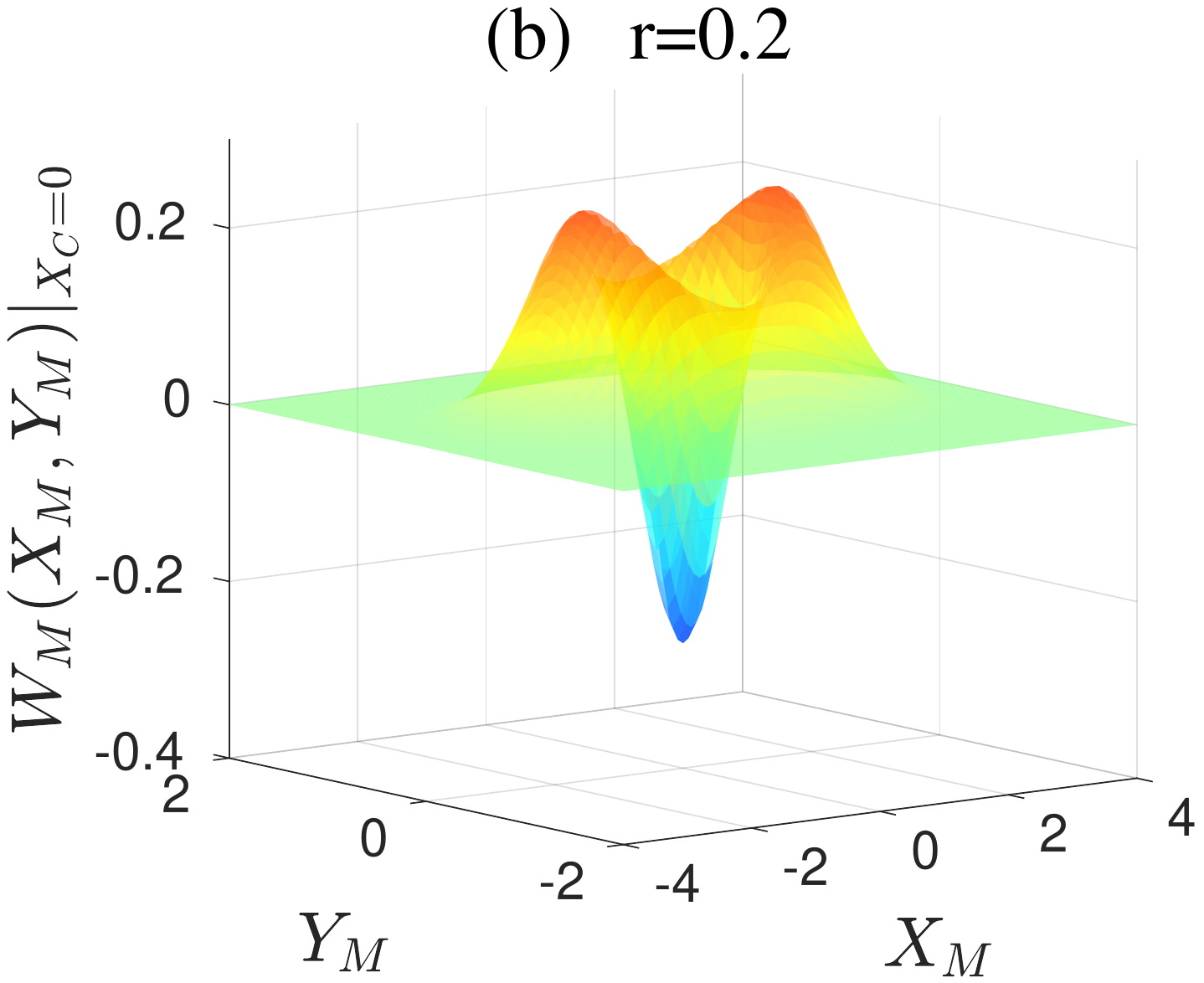}\\
	\includegraphics[width=0.35\textwidth]{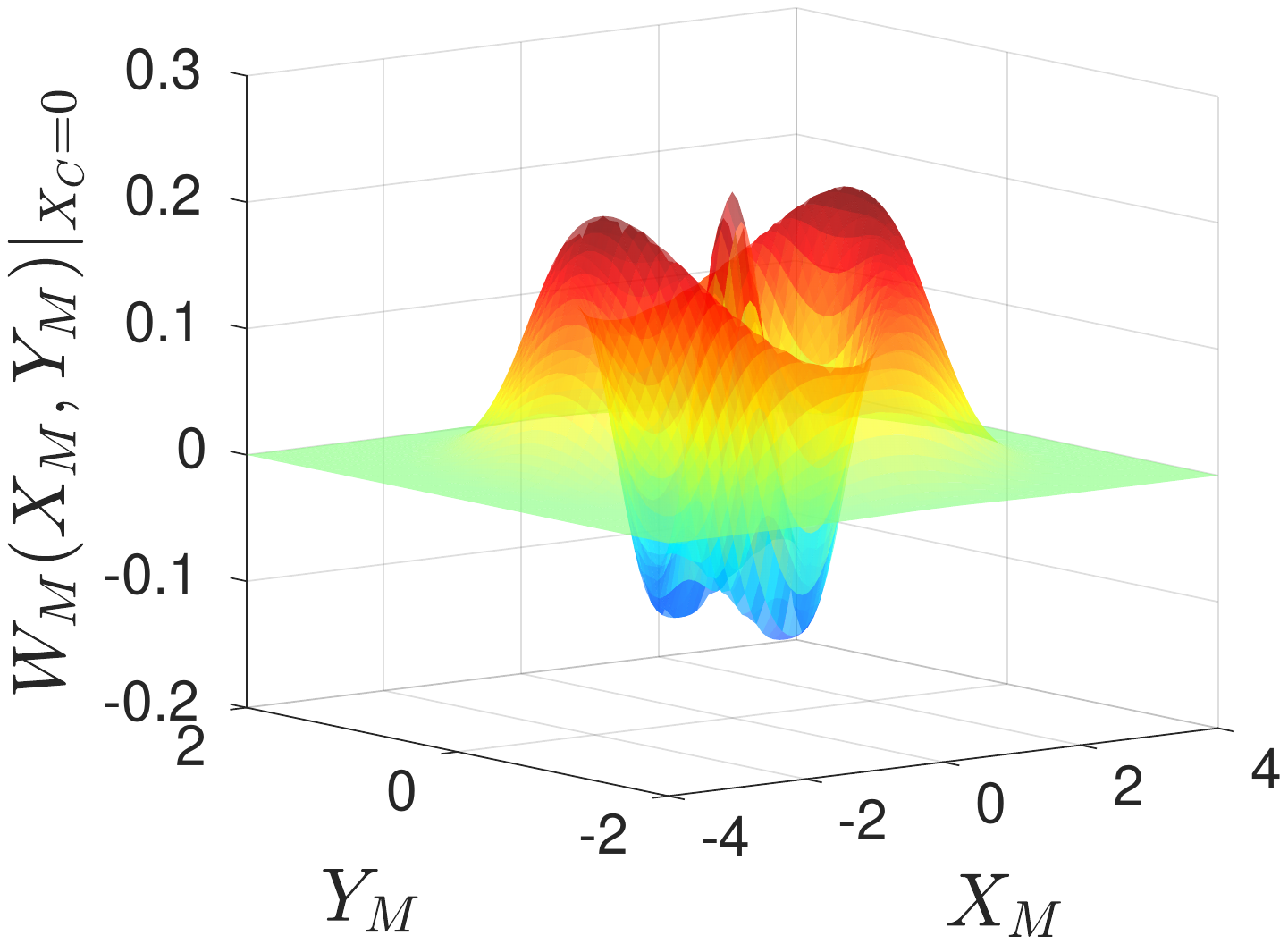}
	\includegraphics[width=0.35\textwidth]{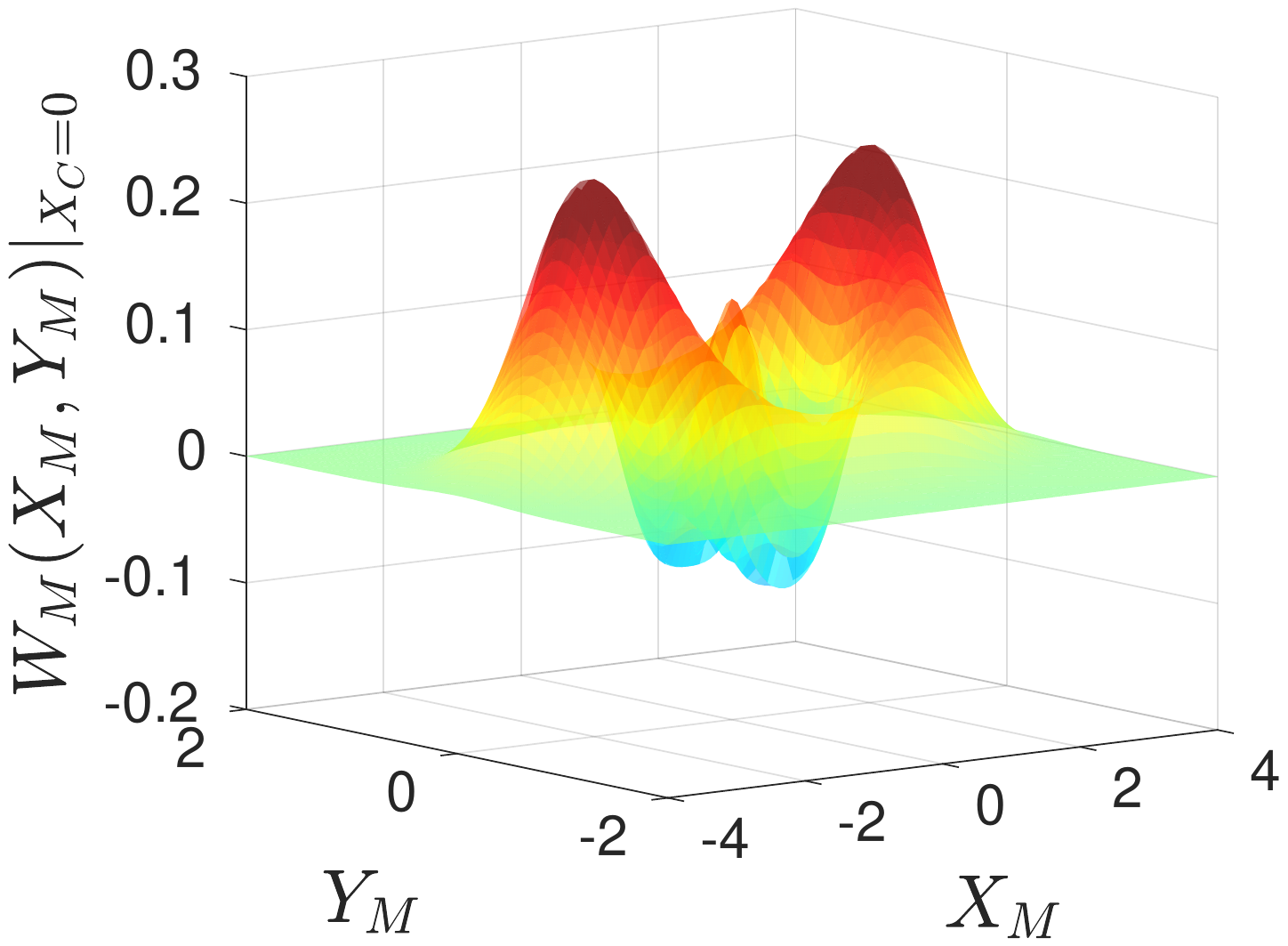}
	\caption{Wigner functions of the transient magnon cat states with temperature  $T=0.3\,$K. The effective squeezing parameters $r=G\tau=0.1$ (a),~$0.2$ (b), where $G=g^2/\kappa-\gamma$ and $\tau$ is the pulse time. Upper row: Wigner function of the projected magnon mode $M$ for the outcome $X_C=0$ after subtracting single photon from the optical mode $C$, displaying features of an odd cat state. Lower row: The transient even cat states by performing a single-photon subtraction and single-photon addition in sequence on the optical mode before further projective measurement. Other parameters are the same as in Fig.~2 of the main text.}
	\label{fig:thermal}
\end{figure}
At the end of this section, we would like to discuss about the effects of thermal occupation of the magnon mode. In Fig.~\ref{fig:thermal}, the magnon mode has been cooled to $T=0.3\,$K, which corresponds to the thermal occupation number $n_{m}=0.2$. It is obvious that the magnon cat states are still able to be generated remotely, while the interference patterns become less apparent than those in Fig.~3 of the main text due to the destructive effects of the thermal occupation. Specifically, the odd (even) magnon cat state at $r=0.2$ is prepared remotely with $|\alpha|^2=1.56\,(3.80)$, fidelity $F=0.66~(0.57)$, Wigner negativity $\delta=0.17~(0.18)$ and macroscopic quantum superposition $I=0.23~(0.24)$. Since the decoherence is accumulated with long interaction time $\tau$, a smaller effective squeezing parameter $r$ will be preferred when considering the thermal occupation. As shown in the Fig.~\ref{fig:thermal}(a), an odd (even) magnon cat state with larger Wigner negativity $\delta=0.23~(0.30)$ and macroscopic quantum superposition $I=0.28~(0.31)$ is generated remotely with $r=0.1$. Consequently, one need to choose appropriate parameters carefully to ensure the quantum qualities required as large as possible.

\section{Effects of the experimental imperfections on the magnon cat states}
        Besides the dissipation resulting from the decay of the magnon and optical modes, the created magnon cat state might be also affected by the direction and the error of the projective measurement, and by the dark counts when performing single-photon operations. In this part, we analyze the effects of above factors, and the lifetime of the cat states created.
\\  
\\
        First, we consider the effect of imperfect projective measurement, including the direction and the probabilistic result of the projection. The results with the measurement $X_C=0$ have been shown in the main text. First, the results of different projective directions of $X_{\theta}=X_{C}\cos\theta+Y_{C}\sin\theta=0$ are demonstrated in the video attached in the Supplemental Materials, namely ``Supplement\_cat\_theta.avi'', whose figures are arranged in a order same as the previous video ``Supplement\_cat\_time.avi''. From the video, we can see that the choice of the projective measurement direction doesn't change the parity and the quality of the cat states, but it just rotates the Wigner function in phase space.
\begin{figure}
	\centering
	\includegraphics[width=0.36\textwidth]{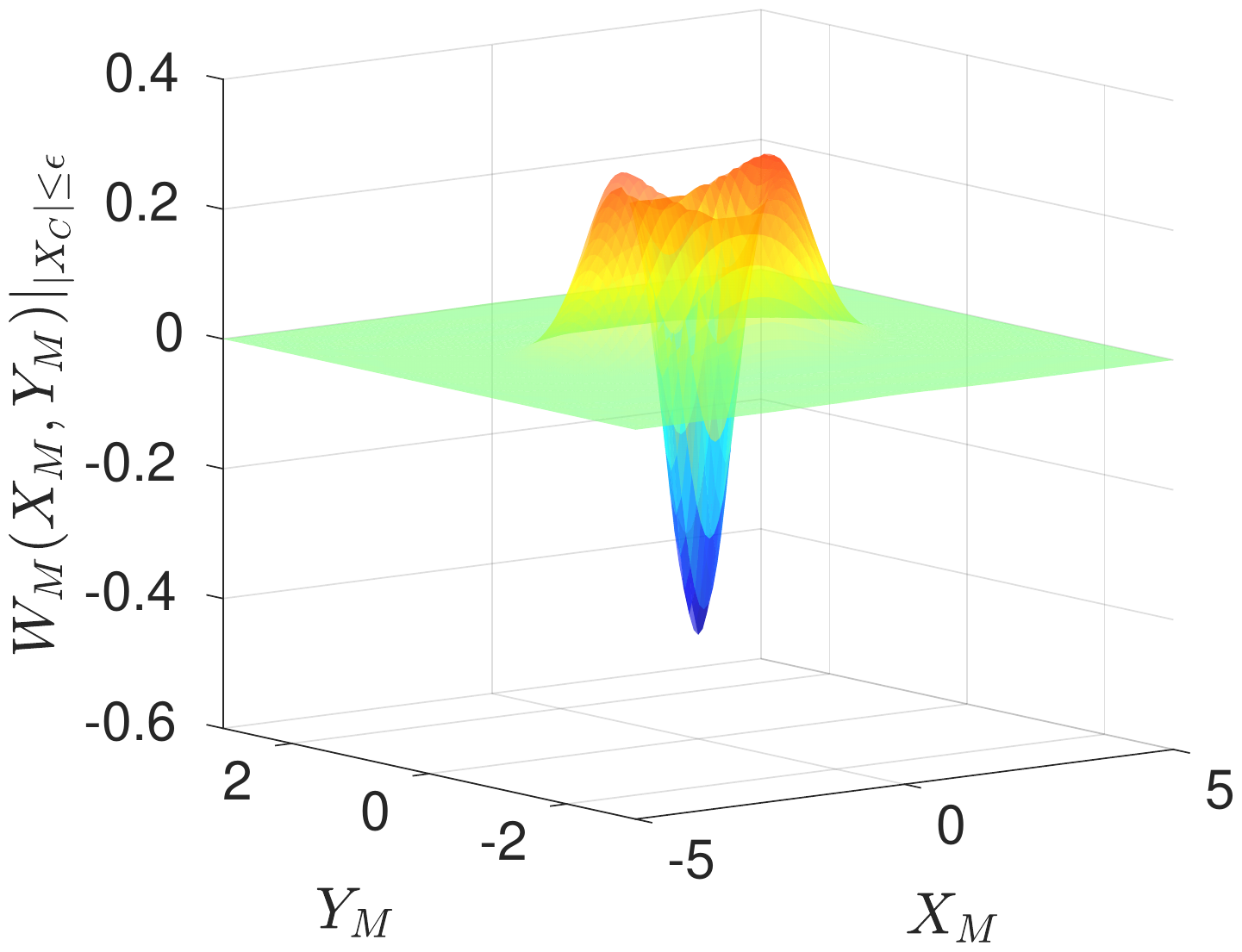}
	\includegraphics[width=0.38\textwidth]{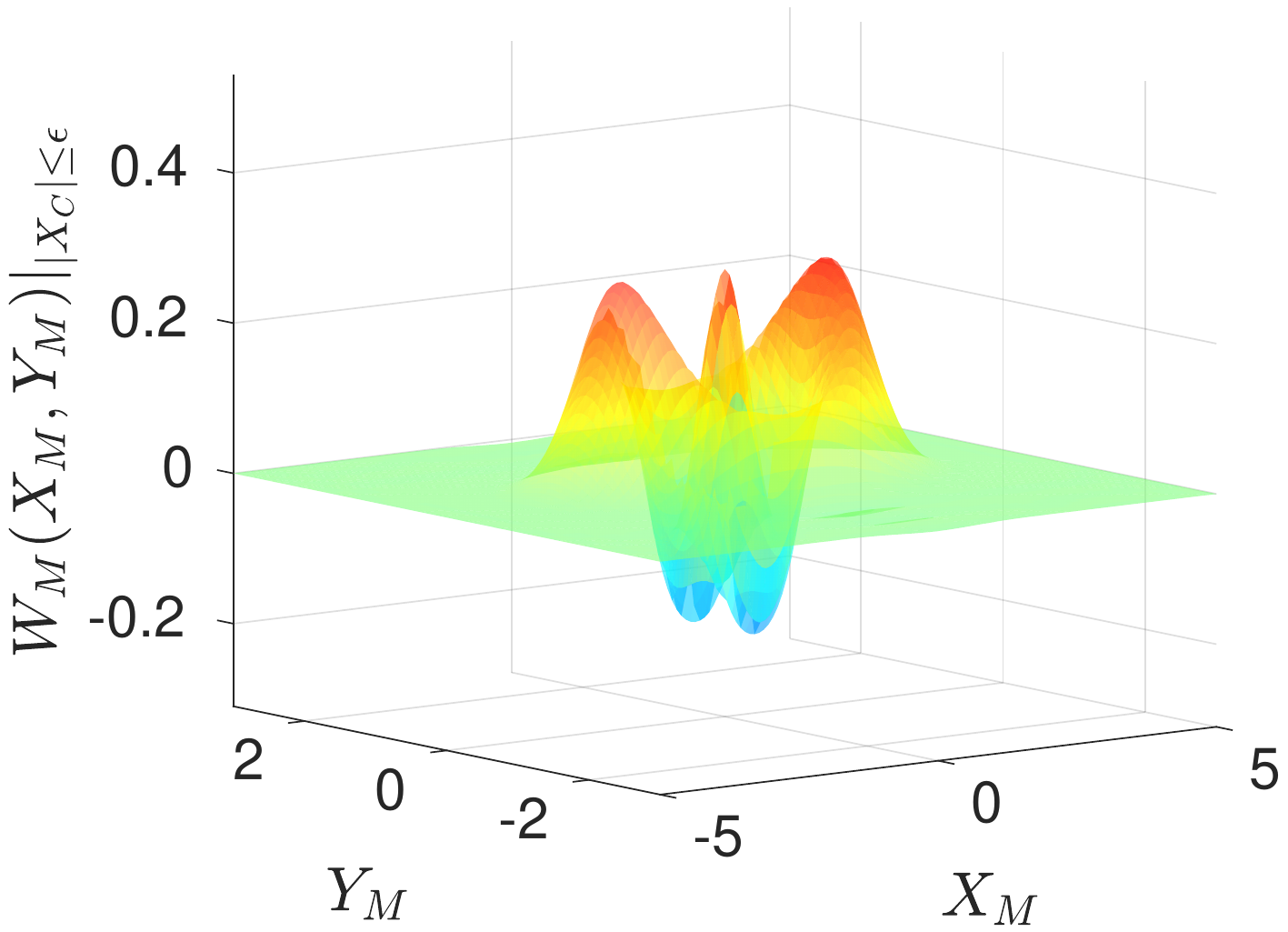}
	\caption{Wigner functions of the transient magnon cat states with a finite projective measurement error $\epsilon=0.1$. The Wigner function of the magnon odd (a) and even (b) cat states for the projective measurement made at $|X_C|\le\epsilon$. The parameters are chosen as $r=G\tau=0.2$, $\gamma=0.1\,$MHz, $g=5\,$MHz, and $\kappa=100\,$MHz, which are the same as Fig.~3(b) in the main text.}
	\label{fig:W-error}
\end{figure}
\\
\\
Taking the experimental imperfection into account, there are always projective measurement errors $\epsilon$ which give us outcome $|X_C|\le\epsilon$ instead of exactly $X_C=0$ precisely~\cite{ourjoumtsev2007generation}. Hence, the fidelity and the quantum coherence of the generated magnon cat states will be affected. We consider here an example where the projective measurement error is $\epsilon=0.1$, and study the effects in Fig.~\ref{fig:W-error}, which corresponds to the case of Fig.~3(b) in the main text. Our results show that the approach we propose is still capable to preparing the magnons in an odd/even cat state considering the imperfection of projective measurements. An odd (even) magnon cat state of $|\alpha|^2=1.44~(3.42)$, fidelity $F=0.84~(0.68)$, Wigner negativity $\delta=0.29~(0.35)$ and macroscopic quantum superposition $I=0.75~(0.80)$ is remotely created. Comparing these results with the ideal projective measurement $X_C=0$ displayed in Fig.~3(b) in the main text, we see that the reduction of quality of the created cat is negligible. 
\\
\\
Second, there always exist dark counts when performing single-photon operations. It is reported that a dark count rate of $n_d\approx10\,$Hz with an overall count rate of $n_o\approx600\,$Hz has been achieved experimentally~\cite{ourjoumtsev2009preparation}, which yields successful probability of the single-photon operations $\xi\approx0.98$. Considering the experimental imperfections of the single-photon operations and the projective measurement simultaneously, the remotely created magnon cat states can be seen in  Fig.~\ref{fig:W-imperfection}. It is observed that an odd/even magnon cat state is generated where the interference patterns become less apparent in such circumstances. To be specific, an odd (even) magnon cat state of $|\alpha|^2=1.44~(3.42)$, fidelity $F=0.81~(0.62)$, Wigner negativity $\delta=0.26~(0.28)$ and macroscopic quantum superposition $I=0.69~(0.71)$ is remotely created. Comparing with the ideal projective measurement $X_C=0$ and successful probability $\xi=1$ displayed in Fig.~3(b) in the main text, we see that the reduction of quality of the created cat is still tolerable. 
\begin{figure}
	\centering
	\includegraphics[width=0.36\textwidth]{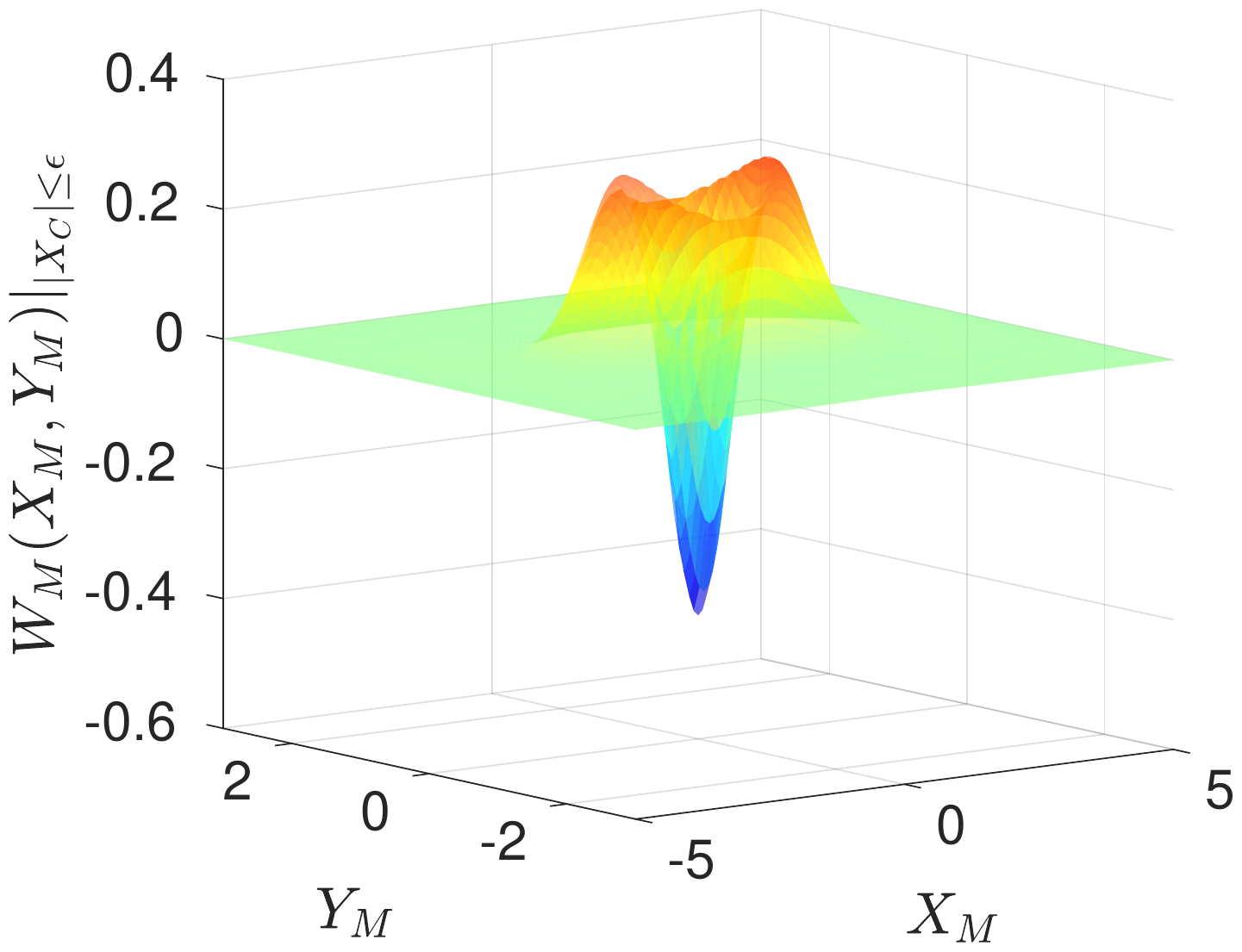}
	\includegraphics[width=0.38\textwidth]{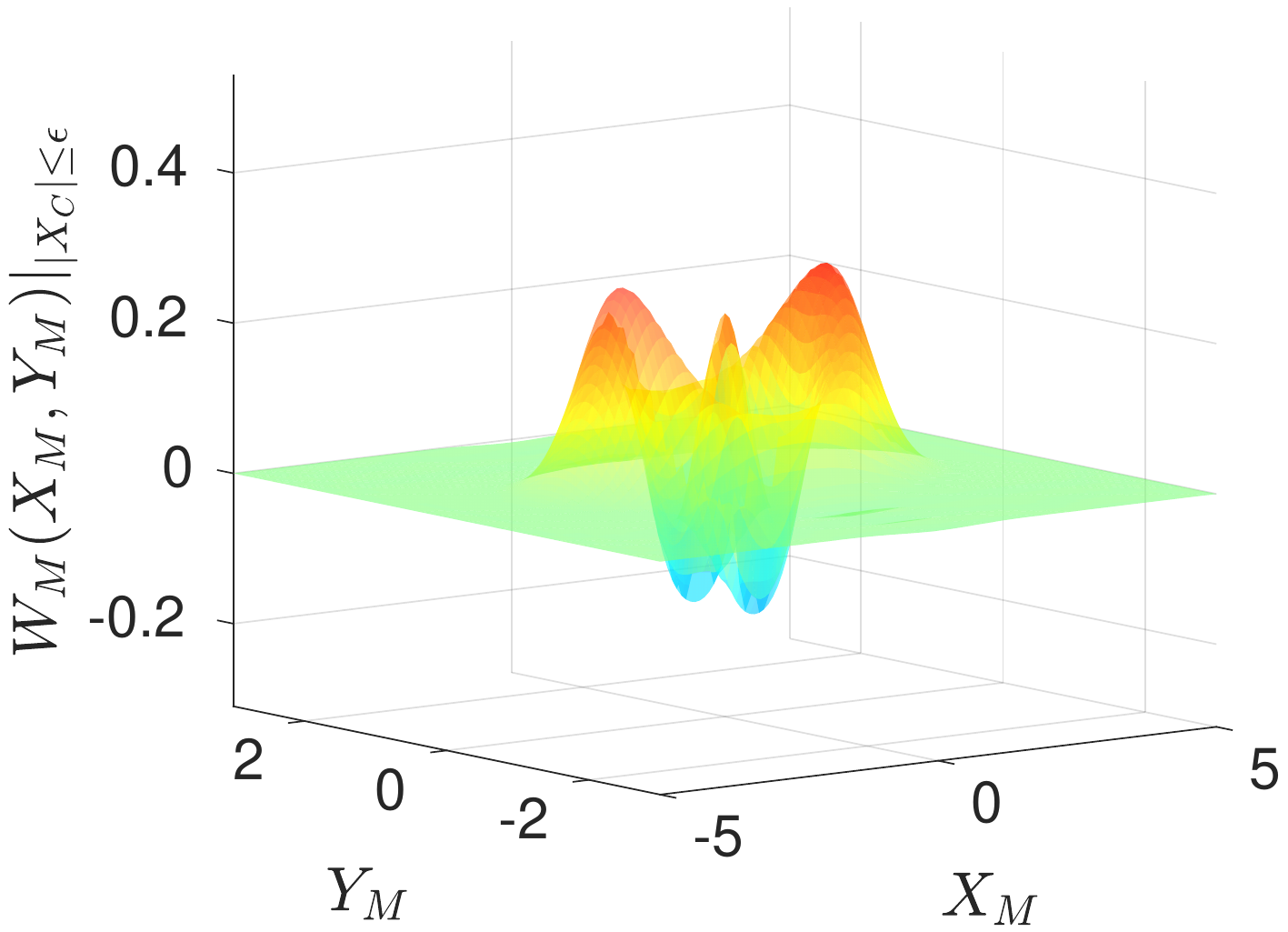}
	\caption{Wigner functions of the transient magnon odd (a) and even (b) cat states with a finite projective measurement error $|X_C|\le\epsilon=0.1$ as well as a successful probability of the single-photon operations $\xi\approx0.98$. The parameters are chosen as $r=G\tau=0.2$, $\gamma=0.1\,$MHz, $g=5\,$MHz, and $\kappa=100\,$MHz, which are the same as Fig.~3(b) in the main text.}
	\label{fig:W-imperfection}
\end{figure}
\begin{figure}
	\centering
	\includegraphics[width=0.345\textwidth]{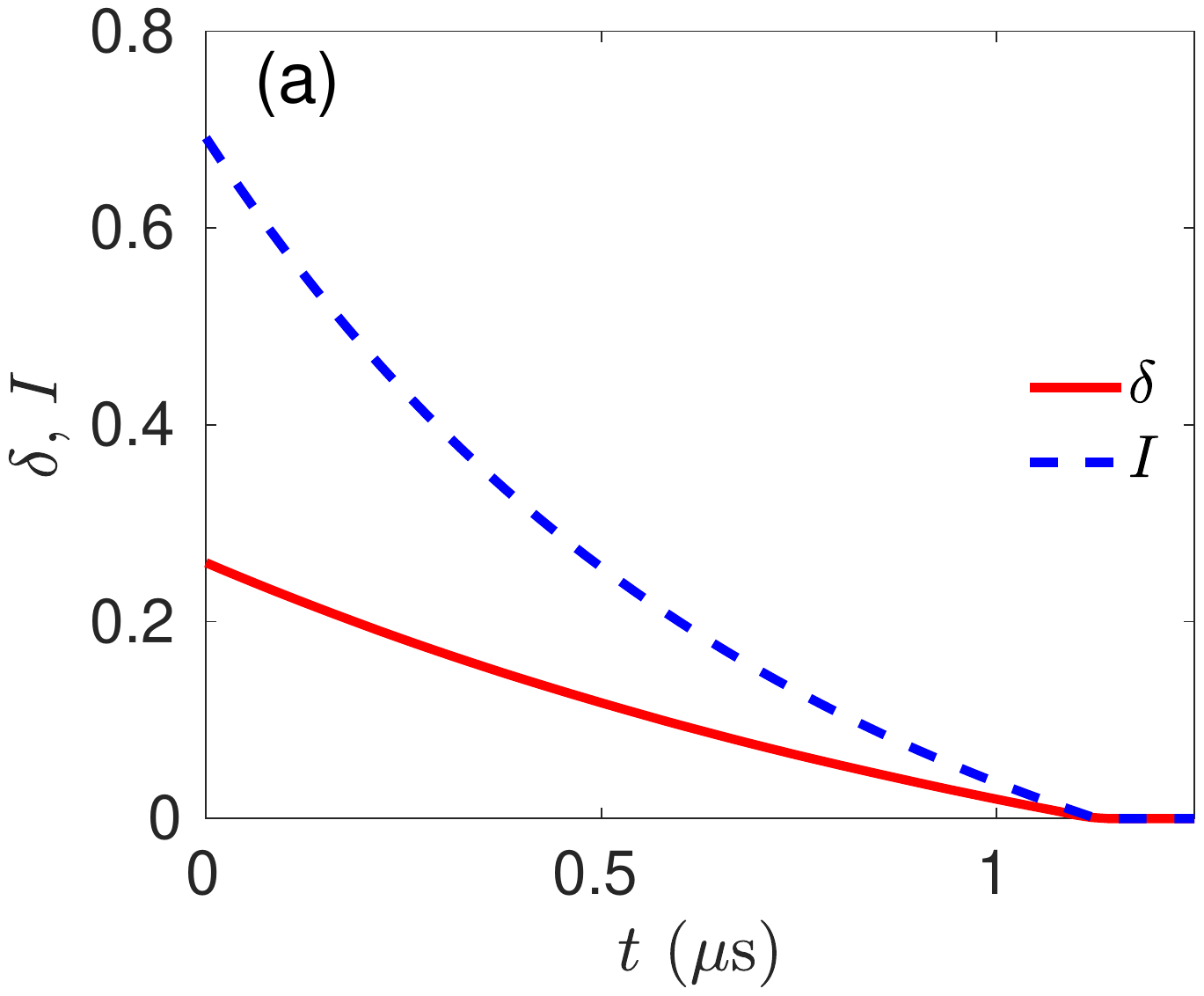}
	\includegraphics[width=0.362\textwidth]{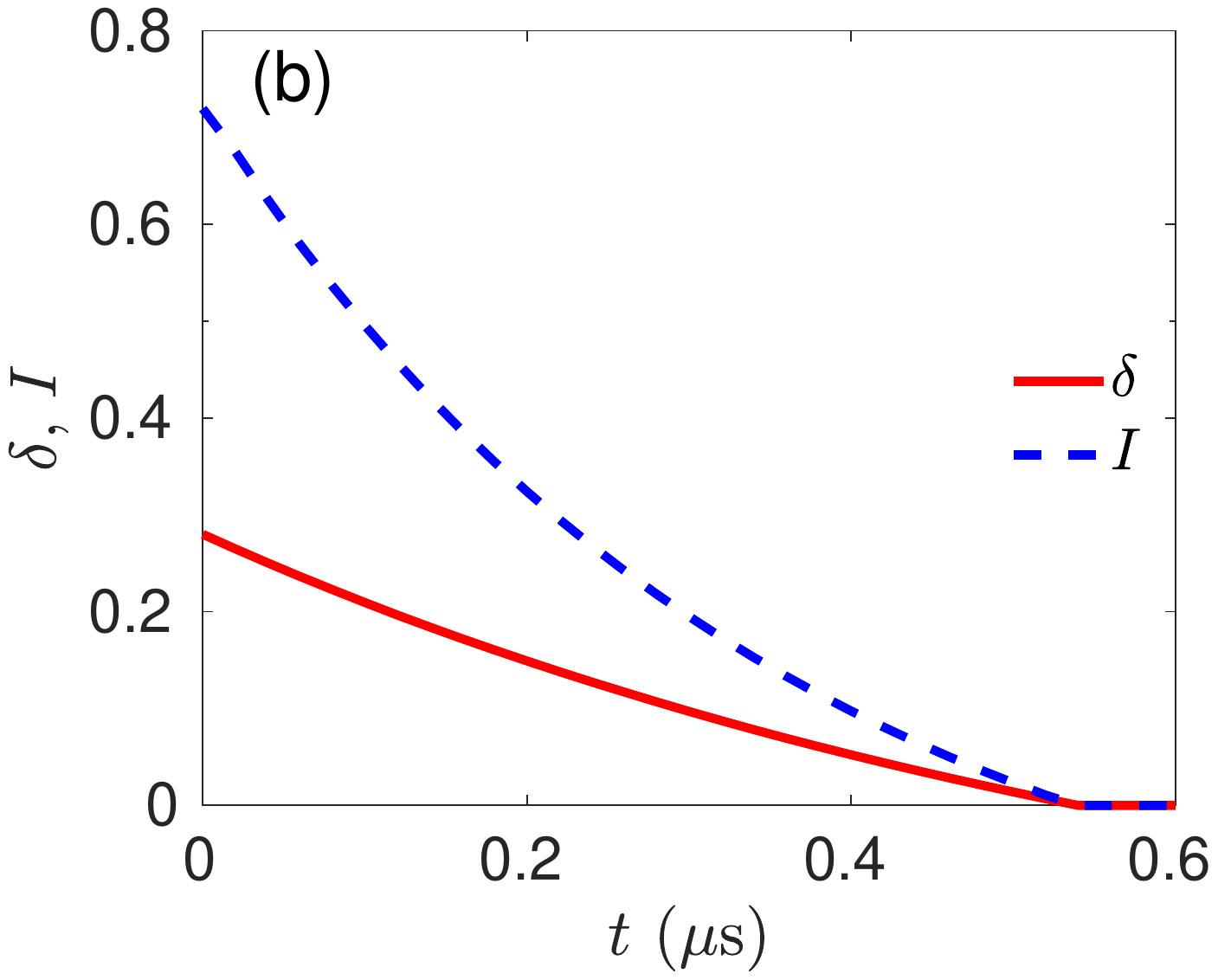}
	\caption{Time evolution of the Wigner negativity $\delta$ (red solid) and macroscopic quantum superposition $I$ (blue dashed) once the magnon odd (a) and even (b) cat states created under the following conditions: $r=G\tau=0.2$, $\gamma=0.1\,$MHz, $g=5\,$MHz, and $\kappa=100\,$MHz, where the projective measurement error $|X_C|\le\epsilon=0.1$ and the successful probability of the single-photon operations $\xi\approx0.98$ have been considered as Fig.~\ref{fig:W-imperfection}.}
	\label{fig:I-time}
\end{figure}
\\
\\
Lastly, we consider the lifetime of the created cat states, which is of crucial importance for concrete applications. After the magnon cat state has been prepared, it will decohere into a classical mixture due to the interaction between the magnon mode and the environment. The evolution of the magnon cat state can be described by the master equation,
\begin{equation}
	\dot{\rho}=-i\omega_m[m^\dagger m,\rho]+\gamma(2m\rho m^\dagger-m^\dagger m\rho-\rho m^\dagger m).
\end{equation}
Here $\rho$ is the density matrix of the magnon mode. In the Wigner representation, the master equation will be transformed to the Fokker-Planck equation, which describes the time evolution of the Wigner function~\cite{scully2012quantum},
\begin{equation}
	\frac{\partial W}{\partial t}=\left[ \left(\gamma+i\omega_m \right)\frac{\partial}{\partial\alpha}\alpha+\left(\gamma-i\omega_m \right)\frac{\partial}{\partial\alpha^*}\alpha^*+\gamma\frac{\partial^2}{\partial\alpha\partial\alpha^*} \right]W.
\end{equation} 
With this method, we examine the time evolution of the Wigner negativity $\delta$ and the macroscopic quantum superposition $I$ for the odd/even magnon cat state displayed in  Fig.~\ref{fig:W-imperfection} where the projective measurement error $|X_C|\le\epsilon=0.1$ and the successful probability of the single-photon operations $\xi\approx0.98$ have been considered. And the results are shown in Fig.~\ref{fig:I-time}. Our results indicate that the lifetimes are $t_{\text{life}}\sim1.12\,\mu$s for the odd magnon cat state and $t_{\text{life}}\sim0.54\,\mu$s for the even magnon cat state. This indicates that the odd magnon cat state survives longer than the even magnon cat state, which is expected because the larger the Schr\"{o}dinger cat is, the more sensitive it becomes to the environment-induced decoherence.

\bibliography{cat-magnon}